\documentclass[twocolumn]{aastex631}
\usepackage{amsmath}
\usepackage{rotating}

\graphicspath{{figures/}}

\usepackage{appendix}

\received{}
\revised{}
\accepted{}

\shorttitle{Truncated PPDs}
\shortauthors{Yang and Bai}

\begin{document}

\title{Global Non-ideal Magnetohydrodynamic Simulations of Protoplanetary Disks with Outer Truncation}

\correspondingauthor{Haifeng Yang, Xue-Ning Bai}
\email{yanghaifeng@tsinghua.edu.cn, xbai@tsinghua.edu.cn}

\author[0000-0002-8537-6669]{Haifeng Yang}
\altaffiliation{C.N. Yang Junior Fellow}
\affil{Institute for Advanced Study, Tsinghua University, Beijing, 100084, China}

\author[0000-0001-6906-9549]{Xue-Ning Bai}
\affil{Institute for Advanced Study, Tsinghua University, Beijing, 100084, China}
\affil{Department of Astronomy, Tsinghua University, Beijing, 100084, China}



\begin{abstract}
It has recently been established that the evolution of protoplanetary disks is primarily driven by magnetized disk winds, requiring large-scale magnetic flux threading the disks. The size of such disks is expected to shrink in time, as opposed to the conventional scenario of viscous expansion. We present the first global 2D non-ideal magnetohydrodynamic (MHD) simulations of protoplanetary disks that are truncated in the outer radius, aiming to understand the interaction of the disk with the interstellar environment, as well as global evolution of the disk and magnetic flux. We find that as the system relaxes, poloidal magnetic field threading the disk beyond the truncation radius collapses towards the midplane, leading to rapid reconnection. This process removes a substantial amount of magnetic flux from the system, and forms closed poloidal magnetic flux loops encircling the outer disk in quasi-steady-state. These magnetic flux loops can drive expansion beyond truncation radius,
corresponding to
substantial mass loss through magnetized disk outflow beyond truncation radius
analogous to 
a combination of viscous spreading and
external photoevaporation. 
The magnetic flux loops gradually shrink over time whose rates depend on level of disk magnetization and external environments, which eventually governs the long-term disk evolution.
\end{abstract}



\section{Introduction} \label{sec:intro}

Global evolution of protoplanetary disks (PPDs) is largely governed by mechanisms that drive angular momentum transport, where magnetic fields are believed to play a crucial role. Conventionally, disk is considered to evolve viscously, which leads to outward angular momentum transport, with turbulent viscosity from the magnetorotational instability (MRI, \citealt{Balbus1991}). In PPDs, however, the MRI is found to be largely suppressed or damped due to the weak level of disk ionization \citep{BaiStone2013b,Bai2013,Simon2013,Gressel2015}, and disk evolution is believed to be primarily driven by a magnetized disk wind, which extracts and carries away disk angular momentum through an outflow mediated by magnetic forces \citep{Blandford1982}.

Launching magnetized disk winds requires large-scale poloidal magnetic flux threading the disks, and the disk winds are launched from disk atmosphere that flows along such poloidal field lines. It is well known that solutions for magnetized disk wind are intrinsically global, determined by both disk microphysics (that determines the coupling of gas with magnetic fields), as well as global boundary conditions (that determine global field configuration). Early studies of disk winds dismissed disk microphysics by treating the disks as razor-thin, and simplified boundary conditions by imposing the self-similar ansatz \citep{Konigl1989,Pudritz1986,Li1995,Krasnopolsky1999}. Later studies relaxed the assumption of razor-think disk and considered disks of finite thickness, usually inserting certain level of disk resistivity to mimic turbulent dissipation 
(e.g., \citealt{Casse2002,Zanni2007,Tzeferacos2009,Sheikhnezami2012}).
More recent studies on PPDs have increasingly incorporated more realistic disk microphysics by taking into account the disk ionization structure, and the resulting non-ideal mangetohydrodynamic (MHD) effects, both in local \citep{BaiStone2013b,Bai2013,Simon2013} and global \citep{Gressel2015,Bethune2017} simulations. Further studies incorporating more realistic thermodynamics reveal that the wind is magneto-thermal in nature \citep{Bai2016}, and mass loss rate from PPD disk winds is likely substantial, comparable to wind-driven accretion rates \citep{Bai2017,WangBG2019,Gressel2020}.

Despite studies that focus on disk microphysics, attention has rarely been paid to global boundary conditions. PPDs have finite size, as set by initial conditions from star formation. In previous global simulations, disks are usually taken to be infinitely extended, usually leading to quasi-self-similar wind structure.
However, as PPD is truncated beyond certain outer radius, the drop in disk density and hence pressure would likely influence the global distribution of magnetic flux and hence the entire wind solution. Consequently, we anticipate that incorporating disk outer truncation would likely alter our views on two most fundamental problems in the theory of disk evolution, stated below.

First, the evolution of the bulk disk mass reservoir. Typically, most of the disk mass is distributed in the outer disk around and beyond truncation radius, and it is such regions that largely sets the disk evolution timescales.
The conventional wisdom is that for viscous disk evolution, the outer disk receives angular momentum from the inner disk and expands, known as viscous spreading (e.g., \citealt{LP1974,Hartmann1998}). On the other hand, for wind-driven accretion, as the disk directly loses angular momentum to the wind, it is anticipated that the disk size should decrease over time. This distinction has been considered in the recent literature to broadly distinguish the two mechanisms by statistical studies of observed disk sizes over age, with tentative evidence in favor of viscous evolution \citep{Tazzari2017,Najita2018,Trapman2020}.\footnote{We also note the complications of tracing disk size from gas (mainly CO, rather than dust mm continuum) observations \citep{Barenfeld2017,Ansdell2018}, which are subject to optical depth effect linked to not-well-understood effect of CO-depletion (e.g., \citealt{Facchini2017,Trapman2019}).} However, as we shall see, incorporating disk truncation complicates the process of wind-driven disk evolution and it is premature to draw conclusions simply based on trends in disk size evolution.

Second, the evolution of poloidal magnetic flux. Typically, the rate of wind-driven disk accretion and mass loss rates directly scales as the amount of magnetic flux threading the disks (e.g., \citealt{BaiStone2013b,Bai2016,Lesur2021}). Therefore, the more fundamental question for global disk evolution is how magnetic flux evolves.
Early works of magnetic flux transport generally belong to the advection-diffusion framework \citep{Lubow1994}, where inward advection of magnetic flux due to viscously-driven accretion competes with outward diffusion from disk (turbulent) resistivity, with later semi-analytical studies that incorporate disk vertical structure \citep{Rothstein2008,Guilet2012,Guilet2013}, and radial resistivity profile \citep{Okuzumi2014,Takeuchi2014}, though
these works largely ignored the wind-driven accretion process and additional non-ideal MHD effects (other than resistivity). More recently, magnetic flux transport is studied from more realistic global disk simulations \citep{BS2017}, semi-analytical local calculations \citep{Leung2019},
and self-similar numerical solutions \citep{Lesur2021}, which generally find outward flux transport whose rates increases with disk magnetization, in addition to other dependencies.
It remains to study, however, how the results are affected by the presence of disk outer truncation, which will likely yield different field geometries with important consequences to global flux transport.

In this paper, we carry out long-term global non-ideal MHD simulations of PPDs with outer truncation. Our simulations are two-dimensional (2D), and incorporate ambipolar diffuison (AD) as the dominant non-ideal MHD effect for outer PPDs (e.g., \citealt{Wardle2007,Bai2011a}). As the first study, and due to the high the computational demand for long-term evolution, we do not aim to accurately capture the key disk microphysics, but instead parameterize the main physical ingredients behind the microphysical processes. It will offer better intuition for interpreting simulation outcomes and dependencies, allowing us to systematically address the two outstanding questions mentioned above.

The structure of this paper is as follows. We describe our 2D global MHD simulations 
of a truncated disk with prescribed ambipolar diffusion and thermodynamics in Section~\ref{sec:method}.
We start with the discussion of the fiducial run and the general evolution picture of the simulations
in Section~\ref{sec:fiducial}, followed by the parameter study focusing on their impacts on the flux 
transport rate in Section~\ref{sec:pars}.
In Section~\ref{sec:discussion}, we discuss our major contributions and caveats.
We summarized and conclude in Section~\ref{sec:conclusion}.

\section{Method}
\label{sec:method}
\subsection{Dynamical Equations}
We use Athena++, a newly developed finite-volume Godunov's scheme MHD code that uses 
constrained transport to enforce divergence free of magnetic fields
\citep{Stone2020}. It is a successor of the widely used Athena MHD code with 
much more flexible coordinate and grid options in
addition to significantly improved performance and scalability. 
We carry out global simulations of PPDs and solve the MHD equations in conservation form:
\begin{equation}
\frac{\partial \rho}{\partial t}+\nabla \cdot (\rho\textbf{v}) = 0,
\label{eq:masscons}
\end{equation}
\begin{equation}
\frac{\partial (\rho \textbf{v})}{\partial t}+\nabla \cdot \left[\rho\textbf{v}\textbf{v}-\textbf{B}\textbf{B}+
\left(P+\frac{B^2}{2}\right)\mathsf{I}\right] = -\nabla \Phi,
\label{eq:mcons}
\end{equation}
\begin{equation}
\frac{\partial E}{\partial t}+\nabla \cdot \left[\left(E+P+\frac{B^2}{2}\right)\textbf{v}-\mathbf{B}(\mathbf{B}\cdot\mathbf{v})\right] = -\Lambda,
\end{equation}
where $\rho$, $\mathbf{v}$, $P$ are gas density, velocity and pressure respectively, $\mathbf{B}$ is
the magnetic field, $\mathbf{E}\equiv P/(\gamma-1)+\rho v^2/2+B^2/2$ with $\gamma$ as the adiabatic index,
$\mathsf{I}$ is the identity matrix, $\Phi=-GM/r$ is the gravitational potential of the protostar,
and $\Lambda$ is the cooling rate. 
Note that in code unites and the above equations, factors of $4\pi$ are absorbed so that magnetic permeability is $1$.
In addition, we have the induction equation with non-ideal MHD 
effects
\begin{equation}
\frac{\partial \mathbf{B}}{\partial t}=\nabla\times(\mathbf{v}\times\mathbf{B})-\nabla\times(\eta_O\mathbf{J}+\eta_A\mathbf{J}_\perp),
\end{equation}
where the $\eta_O$ and $\eta_A$ are Ohmic and ambipolar diffusivities, respectively, 
$\mathbf{b}\equiv \mathbf{B}/B$ is the unit vector for the magnetic field direction,
$\mathbf{J}=\nabla\times \mathbf{B}$ is the current density and
$\mathbf{J}_\perp=-(\mathbf{J}\times\mathbf{b})\times\mathbf{b}$ is the component perpendicular 
to the magnetic field.
Note that we mainly consider AD that dominates the outer disk, and ignore the Hall effect which is more important towards higher-density inner disk regions. We retain Ohmic resistivity for numerical reasons, to be described in Section~\ref{ssec:NI}. 

\subsection{Simulation Setup}

We perform 2D global simulations in spherical-polar coordinates $(r,\theta)$ assuming axisymmetry. In our data analysis, we also use cylindrical radius $R=r\sin\theta$. Our code units are chosen such that $GM=1$, $r=r_0=1$ at the inner boundary. This yields a natural unit of time to be $\Omega_0^{-1}$, where $\Omega_0\equiv\sqrt{GM/r_0^3}$ is the Keplerian frequency at the inner boundary. With these, we define temperature as the ratio of pressure to the density, $T\equiv P/\rho$, with a code unit of $GM/r_0$. We further take the density at disk midplane at the inner boundary $\rho_0=1$.

For the initial condition, we adopt a two-component model: a truncated disk component and a background surrounding 
envelope component, to be elaborated below.

\subsubsection{Disk component}

For the truncated disk component, we adopt a power-law midplane density distribution 
with a Gaussian cut-off:
\begin{equation}
  \rho_{\rm mid}(r) = \rho_0\left(\frac{r}{r_0}\right)^{-\alpha}\exp\left[-\left(\frac{r}{r_c}\right)^2\right],
  \label{eq:rhor}
\end{equation}
where
$r_c$ is the truncation radius, and $\alpha$ is the power-law index of the radial density profile.
The form of density cutoff is not well known observationally, and we choose a Gaussian cutoff is numerically convenient
for its rapid fall-off so that substantial density drop occurs over a reasonable radial range. This profile is also 
subject to later development and will be modified once disk winds are launched. 
The truncation radius $r_c$ is associated with a Keplerian rotation period $P_c=2\pi\sqrt{r_c^3/GM}$, which characterises the dynamical timescale of the outer disk and we will normalize time to $P_c$ when we later discuss long-term evolution.

The temperature profile for the disk component is prescribed with a midplane temperature, $T_{mid}=T_0(r/r_0)^{-1}$, while being constant in cylindrical $z$ direction.
In code unite, we take $T_0=0.01GM/r_0$, which, together with the power law index ${-1}$, 
grants a constant aspect ratio of $H/R=0.1$ for simplicity. 
The corresponding vertical structure of the disk in hydrostatic equilibrium is:
\begin{equation}
\rho_d = \rho_{\rm mid}
\exp\left[\frac{GM}{T_0}\left(\frac{r}{r_0}\right)\left(\frac{1}{r}-\frac{1}{r\sin\theta}\right)\right],
\end{equation}
where $\rho_d$ denote the disk component density.
The initial velocity has zero $r$ and $\theta$ components.
In order to balance the radial pressure gradient and the gravity without magnetic fields in this truncated disk setup, 
the $\phi$ component of velocity takes the following form: 
\begin{equation}
v_\phi^2 = \frac{GM}{r}-(\alpha+1)T_0\frac{r_0}{r\sin\theta}
-2T_0\frac{r_0r\sin\theta}{r_c^2}.
\end{equation}
Note that due to the existence of the third term, $v_\phi^2$ can be negative in some atmosphere regions. 
We set $v_\phi=0$ in such cases.

For our fiducial run, we choose $r_c=30r_0$, corresponding to $P_c\approx1032\Omega_0^{-1}$,
and $\alpha=9/4$,  corresponding to a surface density profile of $\Sigma(r)=\Sigma_0(r/r_0)^{-1.25}\exp\left[{-(r/r_c)^2}\right]$.

The magnetic field in the disk component is initialized to be purely poloidal with constant plasma $\beta_0$ parameter, defined as
$\beta_0\equiv P/P_B$, $P_B=B^2/2$, in the midplane.
To do so, we first calculate the midplane magnetic flux through a numerical integral:
\begin{equation}
\label{eq:Phi_mid}
  \Phi(r) = \int_{r_{B,\rm min}}^{r}  2\pi r'dr'\sqrt{2\rho_{mid}(r)T_{mid}(r)/\beta_0}\ ,
\end{equation}
where $r_{B,\rm min}=0.1r_0$ is the smallest radius in the midplane with penetration of magnetic field lines\footnote{In the midplane, this is inside the disk inner boundary, and field lines originating from this location will enter the simulation domain through the upper/lower parts of the inner boundary.}.
With this definition, the midplane vector potential is given by $A_{\phi,m}(r)=\Phi(r)/(2\pi r)$, and it can be easily
verified that the poloidal field given by $\mathbf{B}=\nabla\times (A_\phi \hat{\phi})$ equals
to $\sqrt{2\rho T/\beta_0}$ at midplane.

The dependence of the vector potential on $\theta$ is taken to be \citep{Zanni2007}:
\begin{equation}
A_\phi(r,\theta) = A_{\phi,m}(r)[1+(m\tan\theta)^{-2}]^{-5/8},
  \label{eq:Azanni}
\end{equation}
where $m$ is a parameter that determines how much the fields bend and we set $m=1$ throughout this work.
For pure poloidal magnetic fields, $A_\phi$ is sufficient to reconstruct the magnetic fields.

Given Equation~\ref{eq:Phi_mid} and the density and temperature profiles, the total magnetic flux in the disk can be 
integrated as:
\begin{equation}
\label{eq:Phid}
\Phi_d = 2\pi r_c^2 \sqrt{\frac{\rho_0T_0}{\beta_0}}\left(\frac{\sqrt{2}r_c}{r_0}\right)^{-(1+\alpha)/2}\Gamma\left(\frac{3-\alpha}{4}\right),
\end{equation}
where $\Gamma(x)$ is the special Gamma function. 
Note that here we have set $r_{B,{\rm min}}=0$ to make the integration analytic. As most magnetic flux resides in the outer disk, the choice of $r_{B,{\rm min}}$ is unimportant for this purpose.
This flux serves as a normalization factor for magnetic flux in our diagnostics later.

\subsubsection{Combination with Envelope component}

\begin{figure}
    \centering
    \includegraphics[width=0.5\textwidth]{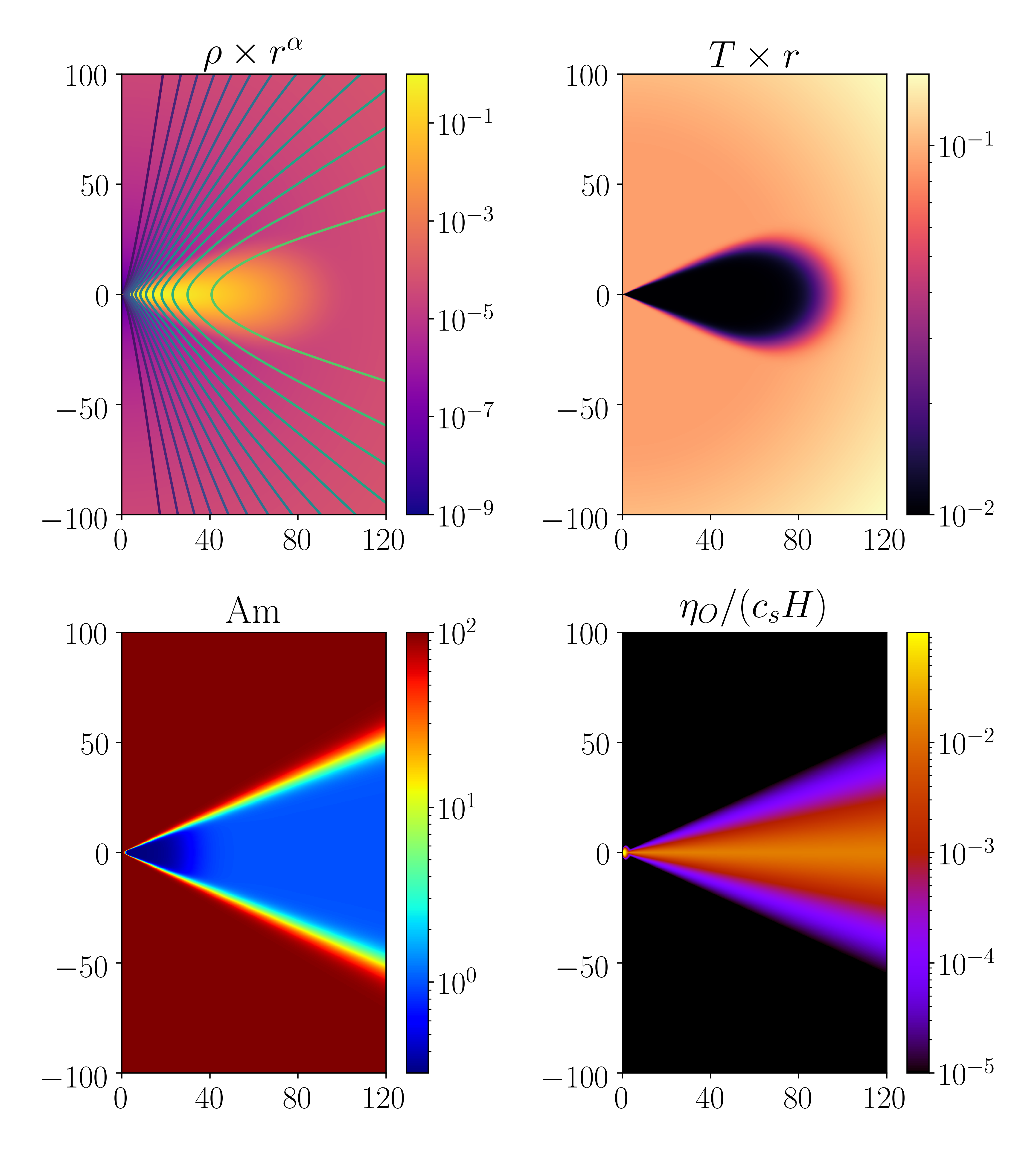}
    \caption{Upper left: the initial density profile for the fiducial run. The poloidal magnetic field lines are overlaid as contours. Upper right: the initial temperature profile for the fiducial run.
    Lower left: the profile of the dimensionless AD Els\"{a}sser number Am for the fiducial run. Lower right: the profile of the artificial reistivity, in unit of $c_sH$, for the fiducial run. }
    \label{fig:am}
\end{figure}

For the envelope component, we adopt a simple power-law model. The density has the same power-law profile 
as the disk component, except for the Gaussian cut-off: $\rho_{e0}(r/r_0)^{-\alpha}$, while being constant in the $\theta$ direction. 
The envelope density at the inner boundary $\rho_{e0}$ is taken as $10^{-9}$ in our fiducial run.
The envelope component is completely at rest, with all three components of velocity being zero throughout the simulation
domain.
The density is summed up to obtain the total density of our model, while the velocities are set as 
the density-weighted values. The initial density profile can be viewed in the upper left panel of Figure~\ref{fig:am}.

The envelope temperature has the same power law index as the disk component at small radii, but with a larger
leading figure, $T_{\rm cor}=T_{c0}(r/r_0)^{-1}$. We call this ``corona temperature" whose meaning will become
obvious soon.
At large radii, the temperature transitions to a constant plataeu value $T_{e0}$:
\begin{equation}
T_e = \left\{
\begin{array}{ll}
T_{c0}\left(\frac{r}{r_0}\right)^{-1}& r\ll r_k\\
T_{e0} & r\gg r_k
\end{array}
\right.,
\end{equation}
where $r_k=r_0(T_{c0}/T_{e0})$ is where these two profiles meet.
These two temperature profiles are connected smoothly through a circular arc in the 
$(\log(r), T)$ plane\footnote{Note that both power-law temperature profile at inner radius and the constant temperature
at large radius are straight lines when plotted in the $(\log(r), T)$ plane. These two lines can be connected through an arc of a circle that is tangent to both of these two lines.
The exact form of transition will be set once the center of the circle is given.
The center of the circle is chosen to be located at $1.1\log(r_k)$ in the $(\log(r),T)$ plane.}. We take $T_{e0}=0.001$ in the fiducial run.

The final temperature of our model is the density-weighted temperature of these two components. At large radii, the temperature is determined by the envelope component.
At small radii, the system transitions from midplane temperature $T_{\rm mid}$ to the corona temperature
$T_{\rm cor} (>T_{\rm mid})$ in polar regions. 
In reality, the atmosphere/corona of the disk is heated to higher temperatures by absorbing the stellar far-UV radiation \citep{Glassgold2004,Walsh2012}.
Since we aim to set up a disk with constant aspect ratio, we enforce the temperature of the disk component
to transition from midplane value $T_{\rm mid}$ to the corona value $T_{\rm cor}$. This transition 
occurs around $\theta = \pi/2 \pm z_\mathrm{trans}(H/r)$, with a transition width of $0.1$\footnote{The transition in this work is prescribed as following. Let $f(x)$ be the function in question, which transitions between two constants
$f_1$ and $f_2$ around $x=x_d$ over width $\delta x$. We then adopt:
$$f(x)=f_1+\frac{1}{2}(f_2-f_1)\left[\mathrm{tanh}\left(\frac{x-x_d}{\delta x/2}\right)+1\right].$$ 
In what follows, we will adopt this formula for transition, unless noted otherwise.
}. 
The far-UV photons heating the corona will also ionize the atmosphere region 
leading to weaker non-ideal MHD effects. Because of this, we also use the same transition height for the AD, which will be discussed in more detail in Section \ref{ssec:NI}.
The transition height is $z_\mathrm{trans}=4$ in the fiducial run. The full initial temperature profile can be viewed in the upper right panel of Figure~\ref{fig:am}.

Note that this combination of disk and envelope components makes the initial disk component not in hydrostatic equilibrium. Without rotation, the envelope component will also tend to fall towards the central object. However, it does not matter much since no equilibrium exists once we introduce magnetic fields, driving accretion and outflows. The prescriptions above simply serve to provide an educated guess for initial density and temperature structure, and the system will relax towards a new steady state. Also, the enevelope component will soon be pushed outside of the simulation domain, as we will see in Section~\ref{sec:fiducial}.

The magnetic field in the envelope is in $z$ direction if converted to a cylindrical coordinate with
a uniform magnitude of $B_{z,e}=3\times 10^{-10}$ in the fiducial run.
This value is chosen such at around $r=100r_0$, the envelope has plasma $\beta\approx 10^5$.
The envelope magnetic field strength is also modified when we change the plasma $\beta_0$ for the disk component to ensure similar plasma $\beta$ in the outer region. 
The vector potential for this uniform magnetic field is simply 
$A_\phi = (1/2)r\sin(\theta)B_{z,e}$.
We have experimented with different choices of the envelope component and the results are largely independent of any specific choice because, as we shall see, the envelope component is quickly blown away after simulation starts.
The vector potential $A_\phi$ of the disk and envelope components are add up to set up the total initial 
magnetic field in our simulation. {The initial poloidal field lines are overlaid in the upper left panel of Figure~\ref{fig:am}.}

As the system evolves, we relax the temperature to the initial temperature at the rate of local Keperian
frequency at the midplane. This corresponds to a cooling rate of:
\begin{equation}
   \Lambda = \frac{\rho}{\gamma-1}\Omega(T_\mathrm{int}-T),
\end{equation}
where $T_\mathrm{int}$ is the initial temperature, and $\Omega=\sqrt{GM/r^3}$ is the local Keplerian angular speed.
We also require the ratio with initial temperature to be no larger than $5$. 

\subsubsection[]{Non-ideal MHD coefficients}
\label{ssec:NI}

For the non-ideal MHD effects, we mainly consider ambipolar diffusion (AD),
which is the dominant effect in the outer regions of PPDs.
In this work the AD diffusivity is prescribed
through the dimensionless AD Els\"{a}sser number $Am\equiv v_A^2/\eta_A\Omega$, with $\Omega$ being 
the disk Keplerian frequency in terms of spherical $r$, and $v_A=B/\sqrt{\rho}$ is the Alv\'{e}n velocity.
We note that unless charged grains become the dominant charge carrier, $\eta_A\propto B^2$ and hence $Am$ is
independent of field strength, and its value is typically on the order of unity in the outer disk \citep{Bai2011b}.
We set $Am=0.3$ in the bulk disk, which gradually transitions to $Am=Am_\mathrm{out}$
beyond $r=r_c$, and to $Am=100$ in the disk atmosphere. The latter ($Am=100$) mimics a strong boost in the
level of ionization due to far-UV (FUV) radiation \citep{PerezBecker2011}, which has become standard
practice in recent disk simulations. 
This transition occurs around $\theta=\pi/2\pm z_\mathrm{trans}(H/r)$ with a transition width of $1(H/r)$. 
This transition of Am resembles that in the temperature profile given their similar physical origins.
The value of $Am_{\rm out}$ is more uncertain, and likely depends on the interstellar environment, and
we choose $Am_{\rm out}=1$ as fiducial value. 
In practice, the $Am$ transition from its inner value to $Am_\mathrm{out}$ around the transition radius $r_c$, and
the transition width is $20r_0$. The Am profile can be viewed in the lower left panel of Figure~\ref{fig:am}.

Besides AD, we also artificially introduces Ohmic resistivity near the inner boundary as well as the midplane
region which solely serves the purpose to better stabilize
the simulations (e.g., \citealp{CuiBai2020}).
In the midplane region, adding such artificial resistivity helps suppress the MRI turbulence and stabilize the current sheet that inevitably develops where $B_\phi$ changes sign\footnote{It is not needed in 3D which can properly capture the MRI turbulence (e.g., see \citealp{CuiBai2021}, which we will consider in the future.}.
The midplane resistivity is prescribed as $\eta_O=f_{\eta,M}c_sH$, with $c_s$ being the sound speed. We take $f_{\eta,M}=0.02$ as fiducial value.
It gradually transitions to $0$ around $\theta\sim \pi/2\pm 0.1$. We have tested that the exact strength of
the artificial resistivity has virtually no effect on our results (see Figure~\ref{fig:phimax}).
We also introduced artificial resistivity near the inner boundary within $1.5r_0$, which is on order of
$0.1c_sH$. The strength of 
the artificial resistivity, in unit of $c_sH$, can be viewed in the lower right panel of Figure~\ref{fig:am}.

\subsubsection[]{Grid setting and boundary conditions}

Because we run simulations over very long time, it generally leads to substantial changes to the
overall surface density profile, which may destabilize regions near the disk inner boundary.
We consider our simulations as numerical experiments under well-controlled
conditions, and such changes would undermine the nature of our simulations. To remedy its negative influence,
we relax the density within the truncation radius $r<r_c$ to the initial value over a timescale of 200 times
the local Keperian period. This approach has little influence to the overall gas dynamics, but allows us to
smoothly run the simulations for dozens of orbits at $r=r_c$ to properly diagnose the results.

We use reflecting boundary conditions in the $\theta$ direction and there is a $2$ degree cone near both poles.
For the inner $r$ boundary condition, we set the hydrodynamic variables in the ghost zones to our initial condition.
In addition, the $v_\phi$ is capped by a rigid body rotation which has the same velocity at $r_0$ with a 
Keplerian rotation. At the outer boundary in $r$ direction, the density is extrapolated assuming a power law index 
$-\alpha$, the same as the initial density profile. The temperature is extrapolated as $T\sim r^{-1}$. 
The $v_r$ and $v_\theta$ are copied to ghost zones but only allows outflow ($v_r\ge 0$). 
The $v_\phi$ is extrapolated according to Keplerian rotation profile.
The magnetic fields are extrapolated with $B_r\sim r^{-2}$, $B_\theta\sim r^0$, and $B_\phi\sim r^{-1}$, for 
both the inner and outer boundary conditions.

The fiducial run has $448$ cells along $r$ direction with logrithmic spacing, and $160$ cells along $\theta$
direction. The $\theta$ grid is designed such that the ratio between adjacent cells is constant, with decreasing cell
size towards the midplane, where the cell size match with the radial grid. 
The ratio between the sizes of adjacent cells in $\theta$ direction 
is $1.01$ in the fiducial run.
This grid gives about $7.8$ cells per scale height in the fiducial run and most other runs.

\subsection{Simulation runs}

We list all our simulation runs in Table~\ref{tbl:runs}. 
Our fiducial run is labeled \texttt{Fid}.
Because we have introduced artificial resistivity, we test whether the results are sensitive to it by conducting two more runs \texttt{etaM1} and \texttt{etaM4}, which have half and twice the
fiducial midplane resistivity, respectively.
Run \texttt{hi\_res} doubles the resolution for convergence study. In
Run \texttt{Rc10} and \texttt{Rc20}, we change $r_c$ to $10r_0$ and $20r_0$, (and $r_\mathrm{max}$ to 100 and 200, accordingly), to examine how our simulation results scales with outer truncation radius. 
To test the impact of the envelope component, we set up Run \texttt{dblRhoe}/\texttt{hlfRhoe} with double/half the 
fiducial envelope density, and Run \texttt{Te1e2}/\texttt{Te2e2} with higher envelope temperature 
($T_{e0}=0.01$ and $0.02$) to make it gravitationally unbound within our simulation domain.
Run \texttt{alpha1.5}, \texttt{alpha2.0} and \texttt{alpha2.5} have different power-law index $\alpha$ 
for the density profile of the disk component, while Run \texttt{HoR05} and \texttt{HoR15} modifies the $H/R$ of the disk component by changing its midplane temperature. 
To examine the impact of the ionization level in external environment, we set up 
Run \texttt{AMout0.3}, \texttt{AMout10} and \texttt{AMout100} with 
$\mathrm{Am}_\mathrm{out}=0.3,\, 10,$ and $100$, respectively.
Through Run \texttt{zt3.5} and \texttt{zt4.5}, we study the role of transition heights in the disk for temperature and AD, with $z_\mathrm{trans}=3.5$ and $4.5$, respectively. 
Finally, we modify disk magnetization in Run \texttt{beta3} and \texttt{beta5} by changing 
the initial poloidal plasma $\beta_0$ to $10^3$ and $10^5$, respectively. The envelope magnetic field $B_{z,e}$ is also changed in proportion.

\begin{table*}
    \begin{center}
    \begin{tabular}{c|cccccccccc|c}
        \hline
        Run & Resolution & $f_{\eta,M}$ & $\beta_0$ & $r_c,r_\mathrm{max}$ & $z_{\rm trans}$ & $\alpha$ & $\rm Am_\mathrm{out}$ & $\rho_{e}$ & $T_{e0}$& $H/R$ & $d\Phi_\mathrm{max}/dt$ \\
        \hline
        \hline   
        Fid      & $448\times 160$ & $0.02$  & $10^4$ & $30,300$ & $4.0$ & $2.25$ & $1.0$ & $10^{-9}$         & 0.001 & 0.1  & -0.0034 \\
        \hline                                                                                                                  
        hi\_res  & $896\times 320$ & $0.02$  & $10^4$ & $30,300$ & $4.0$ & $2.25$ & $1.0$ & $10^{-9}$         & 0.001 & 0.1  & -0.0038 \\
        \hline                                                                                                                  
        etaM1    & $448\times 160$ & $0.01$  & $10^4$ & $30,300$ & $4.0$ & $2.25$ & $1.0$ & $10^{-9}$         & 0.001 & 0.1  & -0.0026 \\
        etaM4    & $448\times 160$ & $0.04$  & $10^4$ & $30,300$ & $4.0$ & $2.25$ & $1.0$ & $10^{-9}$         & 0.001 & 0.1  & -0.0042 \\
        \hline                                                                                                                  
        Te1e2    & $448\times 160$ & $0.02$  & $10^4$ & $30,300$ & $4.0$ & $2.25$ & $1.0$ & $10^{-9}$         & 0.01  & 0.1  & -0.0030 \\
        Te2e2    & $448\times 160$ & $0.02$  & $10^4$ & $30,300$ & $4.0$ & $2.25$ & $1.0$ & $10^{-9}$         & 0.02  & 0.1  & -0.0028 \\
        \hline                                                                                                                  
        dblRhoe  & $448\times 160$ & $0.02$  & $10^4$ & $30,300$ & $4.0$ & $2.25$ & $1.0$ & $2\times10^{-9}$  & 0.001 & 0.1  & -0.0033 \\
        hlfRhoe  & $448\times 160$ & $0.02$  & $10^4$ & $30,300$ & $4.0$ & $2.25$ & $1.0$ & $5\times10^{-10}$ & 0.001 & 0.1  & -0.0036 \\
        \hline                                                                                                                  
        Rc10     & $448\times 160$ & $0.02$  & $10^4$ & $10,100$ & $4.0$ & $2.25$ & $1.0$ & $10^{-9}$         & 0.001 & 0.1  & -0.0035 \\
        Rc20     & $448\times 160$ & $0.02$  & $10^4$ & $20,200$ & $4.0$ & $2.25$ & $1.0$ & $10^{-9}$         & 0.001 & 0.1  & -0.0036 \\
        \hline                                                                                                                  
        alpha1.5 & $448\times 160$ & $0.02$  & $10^4$ & $30,300$ & $4.0$ & $1.5$  & $1.0$ & $10^{-9}$         & 0.001 & 0.1  & -0.0027 \\
        alpha2.0 & $448\times 160$ & $0.02$  & $10^4$ & $30,300$ & $4.0$ & $2.0$  & $1.0$ & $10^{-9}$         & 0.001 & 0.1  & -0.0028 \\
        alpha2.5 & $448\times 160$ & $0.02$  & $10^4$ & $30,300$ & $4.0$ & $2.5$  & $1.0$ & $10^{-9}$         & 0.001 & 0.1  & -0.0057 \\
        \hline                                                                                                                  
        AMout0.3 & $448\times 160$ & $0.02$  & $10^4$ & $30,300$ & $4.0$ & $2.25$ & $0.3$ & $10^{-9}$         & 0.001 & 0.1  & -0.0073 \\
        AMout10  & $448\times 160$ & $0.02$  & $10^4$ & $30,300$ & $4.0$ & $2.25$ & $10$  & $10^{-9}$         & 0.001 & 0.1  & -0.0022 \\
        AMout100 & $448\times 160$ & $0.02$  & $10^4$ & $30,300$ & $4.0$ & $2.25$ & $100$ & $10^{-9}$         & 0.001 & 0.1  & -0.0025 \\
        \hline                                                                                                                  
        zt3.5    & $448\times 160$ & $0.02$  & $10^4$ & $30,300$ & $3.5$ & $2.25$ & $1.0$ & $10^{-9}$         & 0.001 & 0.1  & -0.0089 \\
        zt4.5    & $448\times 160$ & $0.02$  & $10^4$ & $30,300$ & $4.5$ & $2.25$ & $1.0$ & $10^{-9}$         & 0.001 & 0.1  & -0.0018 \\
        \hline                                                                                                                  
        beta3    & $448\times 160$ & $0.02$  & $10^3$ & $30,300$ & $4.0$ & $2.25$ & $1.0$ & $10^{-9}$         & 0.001 & 0.1  & -0.030  \\
        beta5    & $448\times 160$ & $0.02$  & $10^5$ & $30,300$ & $4.0$ & $2.25$ & $1.0$ & $10^{-9}$         & 0.001 & 0.1  & -0.0022 \\
        \hline                                                                                                                  
        HoR05    & $448\times 160$ & $0.02$  & $10^4$ & $30,300$ & $4.0$ & $2.25$ & $1.0$ & $10^{-9}$         & 0.001 & 0.05 & -0.0045  \\
        HoR15    & $448\times 160$ & $0.02$  & $10^4$ & $30,300$ & $4.0$ & $2.25$ & $1.0$ & $10^{-9}$         & 0.001 & 0.15 & -0.0085  \\
        \hline
    \end{tabular}
    \end{center}
    \caption{List of Simulation Runs. In the above table, $f_{\eta,M}$ is the parameter to control the strength of the artificial resistivity in the midplane; $\beta_0$ is the initial plasma $\beta$ of poloidal field; $r_c$ is the truncation radius; $r_\mathrm{max}$ is the maximum radius in the simulation domain; $z_\mathrm{trans}$ defines the transition height for both Ambipolar diffusion and the temperature; $\alpha$ is the power-law index for the density profile; Am$_\mathrm{out}$ is the AD Els\"{a}sser number at large radius; $\rho_e$ is the density at $r_0$ for the envelope component; $T_{e0}$ is the temperature plateau at large radius for the envelope component. In the last column, we also list the measured rate of flux transport $d\Phi_\mathrm{max}/dt$, in unit of $\Phi_d/P_c$. See Section~\ref{subsec:mdot} and \ref{sec:pars} for more details.}
    \label{tbl:runs}
\end{table*}

For all simulations, we first run without magnetic field to $10\Omega_0^{-1}$ before imposing magnetic field.
All simulations are run to a time of at least 20 times the Keperian period at the $r_c$ (roughly
$2\times10^4\Omega_0^{-1}$ in the fiducial run), allowing us to achieve quasi-steady state.

\subsection[]{Diagnostics}
\label{ssec:diags}
In this section, we present several diagnostics that will aid our interpretation.
The first diagnostic is the mass flow/accretion rate, which is very important in understanding the dynamics in the disk. 
Its dependence can be derived from the momentum conservation equation Equation~\eqref{eq:mcons}.
In cylindrical coordinates, the $\phi$ component of this equation reads:
\begin{equation}
\partial_t(\rho v_\phi) 
+ \frac{1}{R^2}\partial_R(R^2 \mathsf{M}_{R\phi})
+\frac{1}{R}\partial_\phi\mathsf{M}_{\phi\phi}
+\partial_z\mathsf{M}_{z\phi} = -\frac{1}{R}\partial_\phi \Phi\ ,
\label{eq:dMphidt}
\end{equation}
where the total stress tensor is:
\begin{equation*}
\mathsf{M}=\rho \mathbf{v}\mathbf{v} - \mathbf{B}\mathbf{B} + \left(P+\frac{B^2}{2}\right)\mathsf{I}\ .
\end{equation*}
Multiplying the both sides of Equation~\eqref{eq:dMphidt} with $R^2$, taking the azimuthal average and 
ignoring the time derivative since we focus on quasi-steady states, we get:
\begin{equation*}
\frac{\partial}{\partial R}(R^2 \overline{\mathsf{M}_{R\phi}}) + R^2\frac{\partial}{\partial z}\overline{\mathsf{M}_{z\phi}} \approx 0\ ,
\end{equation*}
where $\overline{X}$ denotes the azimuthal average of quantity $X$. Plugging in the total stress tensor, we get:
\begin{equation*}
\frac{\partial}{\partial R}(R^2 \overline{\rho v_R v_\phi} - R^2\overline{B_R B_\phi}) 
+ R^2\frac{\partial}{\partial z}\left(\overline{\rho v_z v_\phi}-\overline{B_zB_\phi}\right) \approx 0\ .
\end{equation*}
Note that in our non-ideal MHD simulations in 2D, the gas is largely laminar, we can 
pull $\overline{v_\phi}$ out of the azimuthal average, so that, e.g., 
$\overline{\rho v_Rv_\phi}\approx\overline{\rho v_R}\cdot\overline{v_\phi}$. In addition, we have 
$\partial_R(R\overline{\rho v_R})+R\partial_z(\overline{\rho v_z})=0$ in steady stage due to mass conservation (Equation~\ref{eq:masscons}). With these, we obtain:
\begin{equation}
\overline{\rho v_R} \approx 
  \frac{\frac{\partial}{\partial R}\left(R^2\overline{B_RB_\phi}\right) + 
        R^2\frac{\partial}{\partial z}\left(\overline{B_zB_\phi}\right)
        - R^2\overline{\rho v_z}\frac{\partial\overline{v_\phi}}{\partial z}}
    {R\overline{v_\phi}+R^2\frac{\partial \overline{v_\phi}}{\partial R}},
\label{eq:mflow}
\end{equation}
In general, the first two terms in the numerator dominate, resulting from magnetic stresses, whereas the third term provides a correction at higher altitude where the outflow mass flux becomes significant.
This result will help us interpret the flow structure seen in our simulations.

Our simulations are scale-free, with mass accretion/mass loss rates measured in code units. For units conversion, consider a typical T Tauri disk with $\sim 0.01 M_\odot$ and truncation radius $r_c\sim90$ AU,
the mass accretion rate in physical units is:
\begin{equation}
\begin{split}
\dot{M}_\mathrm{phys} \approx& 5.1\times 10^{-8} M_\odot/\mathrm{yr} \ 
\left(\frac{\dot{M}_\mathrm{code}}{10^{-4}}\right)
\left(\frac{M_\mathrm{disk,phys}}{0.01M_\odot}\right) \\
&\times \left(\frac{M_\mathrm{star,phys}}{M_\odot}\right)^{1/2}
\left(\frac{R_{c,\mathrm{phys}}}{90\rm\, AU}\right)^{-3/2}.
\end{split}
\end{equation}
In above, subscript ``phys" and ``code" denote quantities in physical and code units, respectively. Note that
this units conversion formula applies mainly to our fiducial run, and other simulations with same resolution and disk aspect ratio. 
Under the same assumption, the temperature conversion is simply $GM/r_0\sim 300 \,(\mathrm{km/s})^2$. The sound speed in the envelope with a temperature of $T_{e0}=0.001$ is roughly $0.5\rm\, km/s$, which corresponds to $\sim 30$ K. 
Units conversion for other runs should be straightforward but not shown here.

Finally, we present the diagnostics for magnetic flux transport. 
The rate of flux transport is set by the $\phi$ component of the electric field as \citep{BS2017}:
\begin{equation}
\begin{split}
\frac{d\Phi}{dt} & = -2\pi RE_\phi \\
& = -2\pi R \left[ (v_{e,R}B_z-v_{e,z}B_R)+\eta_O J_\phi)\right]\ .
\end{split}
\end{equation}
This can be further decomposed as:
\begin{equation}
\begin{split}
\frac{d\Phi}{dt} = -2\pi R &[ (v_RB_z-v_zB_R)
+(v_{d,R}B_z-v_{d,z}B_R) \\
&+\eta_O J_\phi]
\equiv -2\pi R(E_{\phi, \rm I}+E_{\phi, \rm NI})\ ,
\end{split}
\end{equation}
where $E_{\phi,\rm I}=v_RB_z-v_zB_R$ and $E_{\phi, \rm NI}=(v_{d,R}B_z-v_{d,z}B_R)+\eta_O J_\phi$ are contributions from the ``ideal MHD" and ``non-ideal MHD" terms, respectively, $\mathbf{v}_d\equiv \mathbf{v}_e-\mathbf{v}\equiv\mathbf{v}_\mathrm{AD}$ 
is the ion-neutral drift velocity due to the ambipolar diffusion (note that we have ignored Hall effect, so that ions and electrons have the 
same velocity), given by:
\begin{equation}
\mathbf{v}_\mathrm{AD} = \frac{\eta_{A}}{B}\mathbf{J}\times\mathbf{B}=
\frac{(\nabla\times \mathbf{B})\times \mathbf{B}}{\rho \Omega\cdot Am}\ .
\end{equation}
This decomposition will help us interpret the microphysical mechanism of magnetic flux transport.

\section{The fiducial run}
\label{sec:fiducial}

In this section, we focus on the fiducial simulation run and its high-resolution counterpart. The results are diagnosed in great detail, as they are representative and contain all major aspects of the conclusions of this paper.

\subsection{General evolutionary picture}
\label{subsec:general}

\begin{figure*}[!htb]
    \centering
    \includegraphics[width=\textwidth]{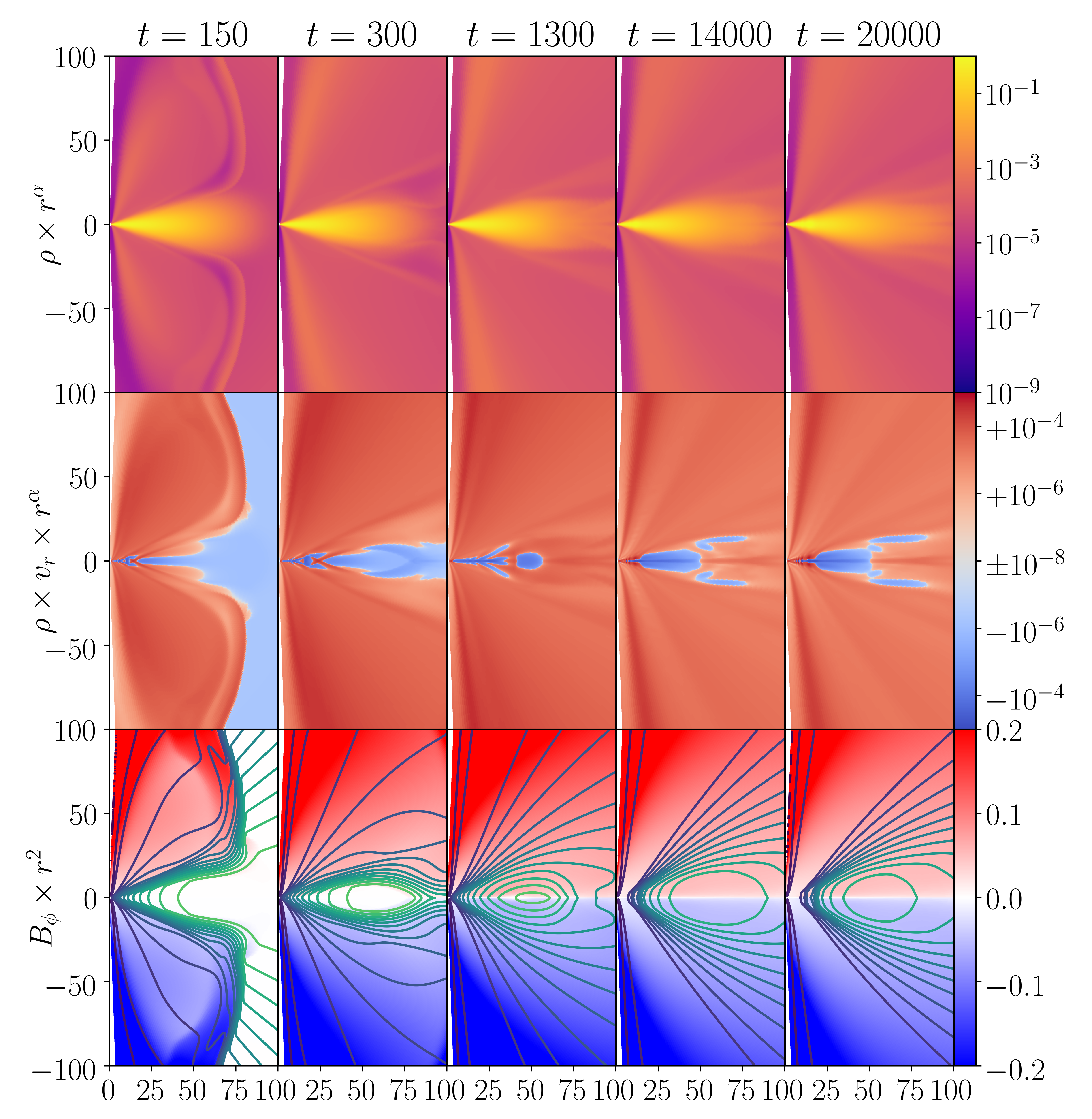}
    \caption{Snapshots of the simulation with fiducial parameters \texttt{hi\_res}. The three rows are, from top to 
    bottom, density $\rho$, radial mass flux $\rho v_r$, and $B_\phi$ with
    magnetic field lines. Factors of some powers of $r$ are applied to enhance
    visibility at large radii. Different columns shows snapshots at different
    time, in unit of $\Omega_0^{-1}$. The poloidal magnetic field lines are shown as contours of constant magnetic flux surface in the third row.}
    \label{fig:snapshots}
\end{figure*}

Following the initial condition shown in Figure \ref{fig:am}, snapshots of subsequent evolution of the fiducial run are shown in Figure~\ref{fig:snapshots}. 
We chose to show only the simulation with high resolution (\texttt{hi\_res}) because the results show no
differences in snapshots, aside from higher resolution. 
Different columns represent snapshots of the simulation at different time. The three rows, from top to bottom,
show the density, mass flow in $r$ direction, and magnetic field structure, respectively. 
The magnetic field lines are shown as contours of constant magnetic flux surface in the third row.
We emphasize that the initial condition is unrealistic and we are primarily interested in the quasi-steady state after evolving the simulations for sufficiently long time where the system largely forgets its initial conditions. Accordingly, the progress of the simulations can be roughly divided into three stages.

At the first stage, magnetic disk winds are launched initially upwards, then radially outwards. 
At the same time, the toroidal component of the magnetic field $B_\phi$ starts to build up
due to the differential rotation in the system. Gradually, the disk wind pushes all the envelope away and dominates the
simulation domain. 
After about one orbital period at the truncation radius $r_c$ ($t\gtrsim1000$), represented by the third column, 
the whole envelope material is pushed away and all the initial magnetic fluxes in the envelope are lost through the outer boundary.
During this process, the material lifted by the disk wind falls down to the midplane carrying magnetic fluxes
with it. This results in magnetic reconnection at the outer region of the simulation domain, which breaks the open field lines and forms loops of poloidal magnetic fields enclosing the outer disk (see first two panels of Figure \ref{fig:snapshots}).

After the rapid relaxation in the first stage, the second stage is a slower relaxation process where we observe the adjustments and fluctuations of density, gas flow and magnetic field structures beyond the truncation radius (not shown in Figure \ref{fig:snapshots}). These variations are still a reflection of initial conditions, which are again not the focus of this study. Towards the end of this stage, corresponding to about 10 orbits at $r_c$, all field loops encircling the outer disk region are smooth and round, with stable flow structure (e.g., see Figures \ref{fig:phimax}, \ref{fig:rt}), marking that we have reached quasi-steady state.

We choose the third, quasi-steady stage to start from $t=14000\Omega_0^{-1}$. During this stage, the magnetic loops gradually shrink and eventually reconnect and disappear from the loop center, which is around $r=50$. It is this stage that likely reflects the reality and will be the focus of our study. While we have run the simulation to much longer time (e.g., see Figure \ref{fig:phimax}), we note that the loss of magnetic flux (through dissipation of poloidal field loops) also changes the dynamics. We thus only consider time intervals after reaching this third stage, while the system still keeps most of its magnetic flux.
For most of the discussions, we average all dynamical quantities from $t=14000\Omega_0^{-1}$
to $t=20000\Omega_0^{-1}$ for diagnostics, unless noted explicitly otherwise. 

There are several important characteristic radii that are important in the diagnostics of the disk dynamics.
In addition to the truncation radius $r_c$, 
there exist a transition radius $R_{t,v}$ in the midplane where the radial
flow changes direction. The net motion of disk material at $R>R_{t,v}$ is outward, while for $R<R_{t,v}$ 
matter is accreting, as can be seen in Figures~\ref{fig:mdot} and \ref{fig:streamline}.
The hydrodynamics is closely related to the poloidal magnetic field structure, which forms loops beyond the 
truncation radius. The center of the loop is defined as $R_{t,\Phi}$, which is the location where the 2D magnetic 
flux function $\Phi(r,\theta)$ is maximized, defined as:
\begin{equation}
\Phi(r,\theta) = \int_0^\theta B_r(r,\theta')\times 2\pi r^2 \sin\theta' d\theta'\ .
\label{eq:Phi}
\end{equation}
We will see that $R_{t,v}$ and $R_{t,\Phi}$ are closely related later in Section~\ref{subsec:mflow}.

Across all three stages in our simulations, we observe a steady transport of magnetic fluxes, angular momentum, and mass in the disk region within the truncation radius $r_c$. If we zoom in to just this part of
the simulation, we find similar results as those discussed in \cite{BS2017}, at qualitative level.
This disk region mainly serves as a sanity check and will not be the focus of our discussion in this paper, but measurements of some fundamental quantities will be presented later in this section for more quantitative comparison.

\subsection{Accretion rate and mass loss}
\label{subsec:mdot}

We start by analyzing accretion and mass loss rates in our fiducial simulation.
The presence of disk outer truncation presents some ambiguity regarding where the disk boundary is when calculating the disk mass loss rates.
In Figure~\ref{fig:mdot}, we show the mean density and velocity vectors
in stage three in the upper left panel,
where the velocity vectors are in unit length to enhance the visibility. 
In this figure, we define a contour to calculate the accretion rate and mass loss quantitatively. This contour
consists of three parts: two constant $\theta$ lines with $\theta=\pi/2\pm \delta_b$ where $\delta_b=3.5(H/r)=0.35$, and one arc with $r=100$.
Here, we define the contour with $3.5(H/r)$ lines instead of $z_\mathrm{trans}(H/r)$, with $z_\mathrm{trans}=4$.
This is partly because our diagnostics, written in cylindrical coordinates, are more accurate for geometrically-thin disks. Also, our transition for temperature and Am has finite transition width of $1(H/r)$, and we find choosing $3.5(H/r)$ above/below the midplane as the ``wind base" helps us better diagnose disk accretion.
As seen in the upper right panel of Figure~\ref{fig:mdot}, and will be clear in later discussions in this section, important flow structures develop well beyond the disk truncation radius, and hence we choose $r=100$ to properly enclose such flow structures. The choice of the arc shape is mainly for simplicity.
We have also tried some other contours and found qualitatively similar results.

The total accretion rate in the disk region is measured by $\dot{M}_\mathrm{acc}=\int_{-z_b}^{z_b}\rho v_R 2\pi R dz$, where $z_b=R\sin\delta_b$, $v_R$ is the radial velocity in cylindrical coordinates. 
The radial profile of $\dot{M}_{\rm acc}$ is shown in Figure \ref{fig:mdot}. We note that the accretion rate plunges within $15r_0$. This is
because as the disk has evolved for very long time, most magnetic flux threading this region has already migrated outwards (see Figure~\ref{fig:snapshots}), leaving little magnetic flux to drive accretion. Nevertheless, this region could also be affected by the inner boundary conditions after long-term evolution and is not the focus of our work.

The mass loss through the disk region, defined as the mass outflow across the contour, is calculated as a function 
of cylindrical radius $R$ as follows:
\begin{equation}
\frac{d\dot{M}}{d\log R} = \frac{2\pi R^2}{\cos\delta_b}\rho v_\theta\ ,
\end{equation}
where the $\cos\delta_b$ accounts for a geometric correction in surface area, as we measure mass loss rate per logarithmic cylindrical radius. This differential mass loss rate is measure up to the truncation radius $r_c$.
We can see from the bottom panel of Figure~\ref{fig:mdot} that the $dM/d\log{R}$ through the disk wind is smaller than the accretion rate by a factor of a few, but the integrated mass loss over radius is usually larger in the wind (by a factor of $2-3$ in the fiducial run). 
We comment that the strong mass loss is a typical outcome when conducting simulations with simplified thermodynamics around the wind launching region (e.g., \citealt{BS2017,CuiBai2020,Rodenkirch2020}). Simulations with more realistic treatment of thermal physics tend to yield milder mass loss rates \citep{WangBG2019,Gressel2020}. We thus do not pursue further discussions on disk mass loss in this work.

\subsubsection[]{Mass loss beyond truncation radius}\label{ssec:mlossrc}

\begin{figure}[!htb]
    \centering
    \includegraphics[width=0.5\textwidth]{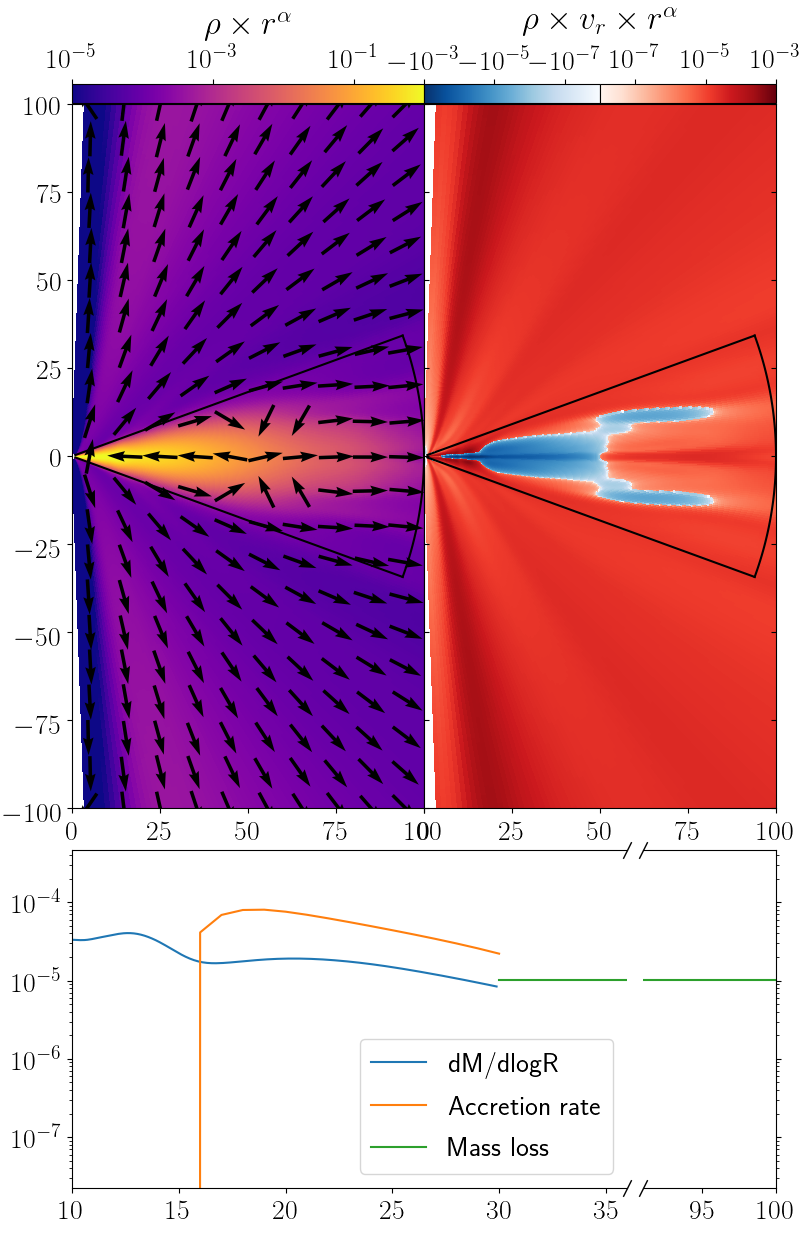}
    \caption{The diagnostics for mass flow, averaged from $14000\Omega_0^{-1}$ and $20000\Omega_0^{-1}$ for 
    the \texttt{hi\_res} run. The upper left panel shows the density, 
    multiplied with $r^\alpha$, over plotted with the unit-length vectors showing poloidal velocity directions.
    The upper right panel is the radial mass flux (red for outflow). The contour in these two panels is defined with two constant $\theta=\pi/2\pm0.35$ lines and an arc at $r=100$.
    The lower panel shows the mean accretion and mass loss rates from the bulk disk within $r_c=30$, and a single integrated mass loss rate from beyond $r_c$. 
    See Section \ref{subsec:mdot} for more details.}
    \label{fig:mdot}
\end{figure}

Beyond the truncation radius, the ``wind base" as well as the bulk disk are not well defined. The magnetic field also starts to form loops, the flow direction becomes more horizontal.
We will discuss the dynamics and mass flow in this region beyond the truncation radius 
more carefully in Section~\ref{subsec:mflow}.
For the purpose of measuring mass loss rates beyond truncation radius, we simply quote a single value, by integrating the mass flux through the remaining parts of the contour (from $r_c$ to radius $100$ and the arc).
The result is shown in the green horizontal line in the bottom panel of Figure~\ref{fig:mdot}. Note that the value that we quote is the net mass loss rate, where only mass loss from the region is counted, whereas mass flux {\it through} the region cancels out.

Again, we see appreciable mass loss rate beyond $r_c$ is comparable to the accretion rates in the bulk disk. Note that such mass loss is usually attributed to external photoevaporation, as a result of (primarily) UV heating from nearby massive stars (e.g. \citealt{Hollenbach1994,Clarke2007}). Despite that our simulations treat the thermodynamics external to the disk in a very rough manner by prescribing the temperature, we emphasize that the temperature that we prescribe is insufficient for the gas beyond disk truncation radius to drive an outflow on its own (at least up to a radius of $\sim1000$ for $T_{e0}=0.001$). Therefore, the outflow seen in our simulations is likely of magnetic origin.

We have conducted some further analysis to probe the origin of the mass loss beyond $r_c$ (but typically within $r_t$), discussed in Appendix \ref{app:mlossrc}, and conclude that about $70\%$ of this mass loss rate can be attributed to the continuation of magnetized disk wind launched beyond $r_c$, which is unbound, 
and about $20\%$ can be attributed to a decretion flow to be discussed in Section \ref{subsec:mflow}. 
These flows can be further seen in the streamline plot in Figure \ref{fig:streamline}, and are smoothly joined at large disk radii.

\subsection{Magnetic flux evolution}
\label{subsec:flux}

\begin{figure}[!htb]
    \centering
    \includegraphics[width=0.5\textwidth]{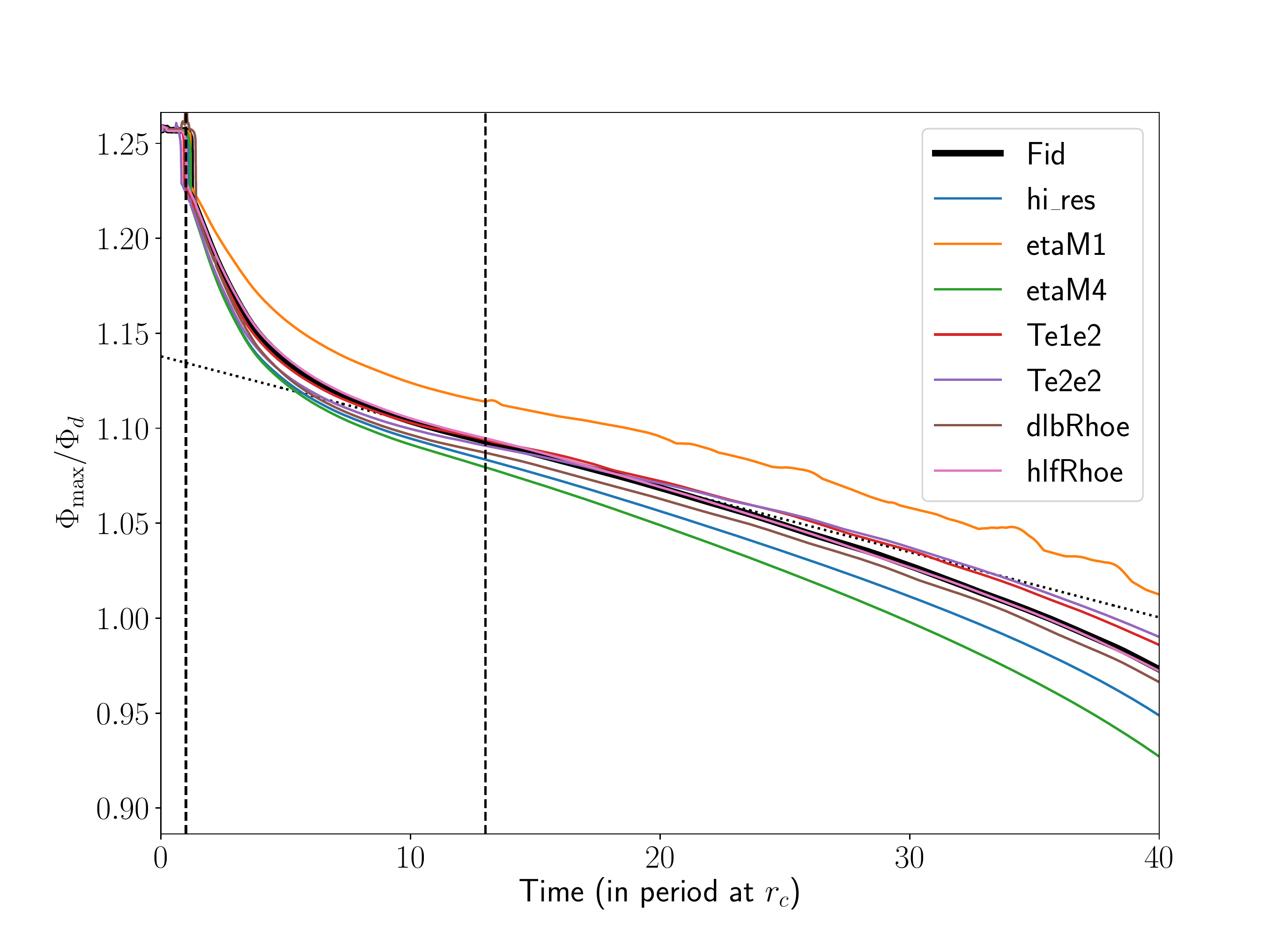}
    \caption{Evolution of maximum magnetic flux $\Phi_\mathrm{max}$ for different runs, where
    $\Phi_\mathrm{max}$ is normalized by $\Phi_d$, the initial magnetic flux threading the disk,
    and time is in unit of the orbital period at the truncation radius $r_c$, $P_c$.
    We fit a straight line of the curve between $10P_c$ and $20P_c$, shown as the dotted line (for \texttt{hi\_res} run).
    The slope of this line represents the rate of magnetic flux loss through reconnection at loop center. The two vertical dashed lines at $t=P_c$ and $t=13P_c$ separate three stages in the simulation, 
    as discussed in Section~\ref{subsec:general}.}
    \label{fig:phimax}
\end{figure}

In the scenario of wind-driven accretion, long-term disk evolution is largely governed by how magnetic flux evolves in disks. We have already seen from simulation snapshots about 
the evolution of the magnetic fluxes. More quantitatively, in Figure~\ref{fig:phimax},
we show the ``flux evolution curve'', which is the maximum value of the magnetic flux
$\Phi$ defined as Equation~\eqref{eq:Phi} across the simulation domain plotted against time.
Note that the time unit here is the orbital period at $r_c$ ($P_c$), and the $\Phi_\mathrm{max}$ is normalized with the total flux in the disk $\Phi_d$, defined in Equation~\eqref{eq:Phid}.

Where the maximum magnetic flux $\Phi_\mathrm{max}$ is achieved depends on the evolutionary stage. At the beginning (the first stage), $\Phi_\mathrm{max}$ is always achieved at the disk outer boundary $r_\mathrm{max}$. As disk wind pushes the envelope away, magnetic fluxes originally threading through the envelope component are quickly lost through the outer boundary. This process is exhibited as a sudden drop in Figure~\ref{fig:phimax}. After all the envelope fluxes are pushed away forming magnetic field loops (second and third stages),
$\Phi_\mathrm{max}$ is always achieved at the loop center, and decreases as the field loop dissipates and shrinks.
Therefore, this ``flux evolution curve'' is a good diagnostic on how fast the flux is lost through this dissipation process.
We see that $\Phi_{\rm max}$ decays over time, and upon entering stage 3, the rate at which flux is lost is largely steady. 
To quantify this flux loss rate in the quasi-steady stage, we fit the curve between $10P_c$ and $20P_c$ with a straight line, shown as the dotted line in Figure~\ref{fig:phimax}. We do not use data beyond about $20P_c$ because the dynamics will be affected by the loss of flux towards later time.
The slope of this line for the fiducial run is $-0.003436 \Phi_d/P_c$, corresponding to a flux loss timescale of $\sim300P_c$.

Also plotted are the results for runs with different artificial resistivity (\texttt{etaM1} and \texttt{etaM4}), 
higher resolution (\texttt{hi\_res}), higher envelope temperature (\texttt{Te1e2} and \texttt{Te2e2}) and different initial
envelope density (\texttt{dlbRhoe} and \texttt{hlfRhoe}). 
For these simulations, the flux evolution curves are generally similar to each other. 
The rate of flux loss ($d\Phi_\mathrm{max}/dt$) is also tabulated in Table~\ref{tbl:runs}.
We verify that the introduction of artificial resistivity does not affect our results much other than stabilizing the simulations.
We also find that the initial parameters in the envelope component, either density or temperature, have little impact on the end result. This is reasonable, since the entire envelope is quickly pushed away during the first stage. Parameters that do affect the loss rate of magnetic flux will be discussed in Section~\ref{sec:pars}.

\subsection{Flow structure beyond truncation radius}
\label{subsec:mflow}

\begin{figure}[!htb]
    \centering
    \includegraphics[width=0.5\textwidth]{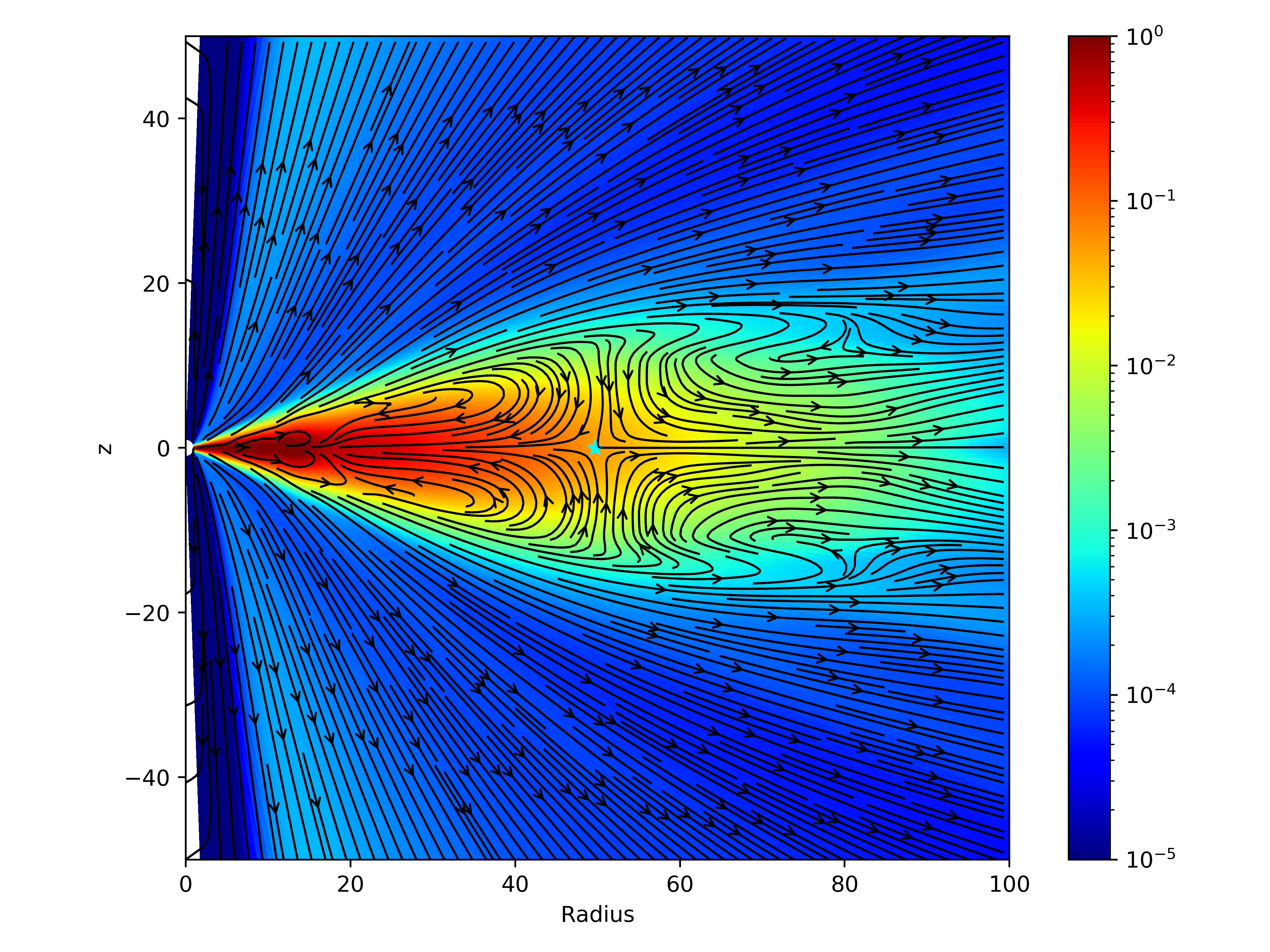}
    \caption{The streamline plot averaged between $14000\Omega_0^{-1}$ and $20000\Omega_0^{-1}$ for Run \texttt{hi\_res}. 
    The colormap represents $\rho r^\alpha$. The streamlines are velocity streamlines. 
    The cyan star at $R_{t,v}=49.7$ marks the location where $v_R$ changes sign in the midplane.
    It largely coincides with $R_{t,\Phi}=49.4$ where the magnetic flux $\Phi$ is maximized and it defines the center of the magnetic flux loops. }
    \label{fig:streamline}
\end{figure}

In this subsection, we focus on the flow structure beyond the disk truncation radius. 
In Figure~\ref{fig:streamline}, we show a streamline plot averaged between $14000\Omega_0^{-1}$ and 
$20000\Omega_0^{-1}$. 
We can see that in the midplane region, there exists a transition point (marked with a cyan star at $R_{t,v}=49.7$). 
Inside this radius, gas flows inward, which is essentially the wind-driven accretion flow. Outside of this radius, however, the gas flows outward, indicative of a decretion flow. In other words, the outer disk expands. This is contrary to the conventional wisdom that the entire disk shrinks under wind-driven disk evolution scenario, as we study in further detail below.

This transition point closely follows the location of the loop center (at $R_{t,\Phi}=49.4$), as we shall see more clearly in Figure \ref{fig:rt}. The reason can be qualitative understood.
We see that toroidal field configuration is largely unchanged within and beyond the loop center. On the other hand, vertical field changes sign across the loop center.
As a result, the Maxwell stress exerted to the disk, $B_zB_\phi$ (more specifically, the sign of the Lorentz force), changes sign when moving beyond the loop center. This sign change leads to a reversal of the magnetic torque, and hence a transition from accretion to decretion flow.

\begin{figure}[!htb]
    \centering
    \includegraphics[width=0.5\textwidth]{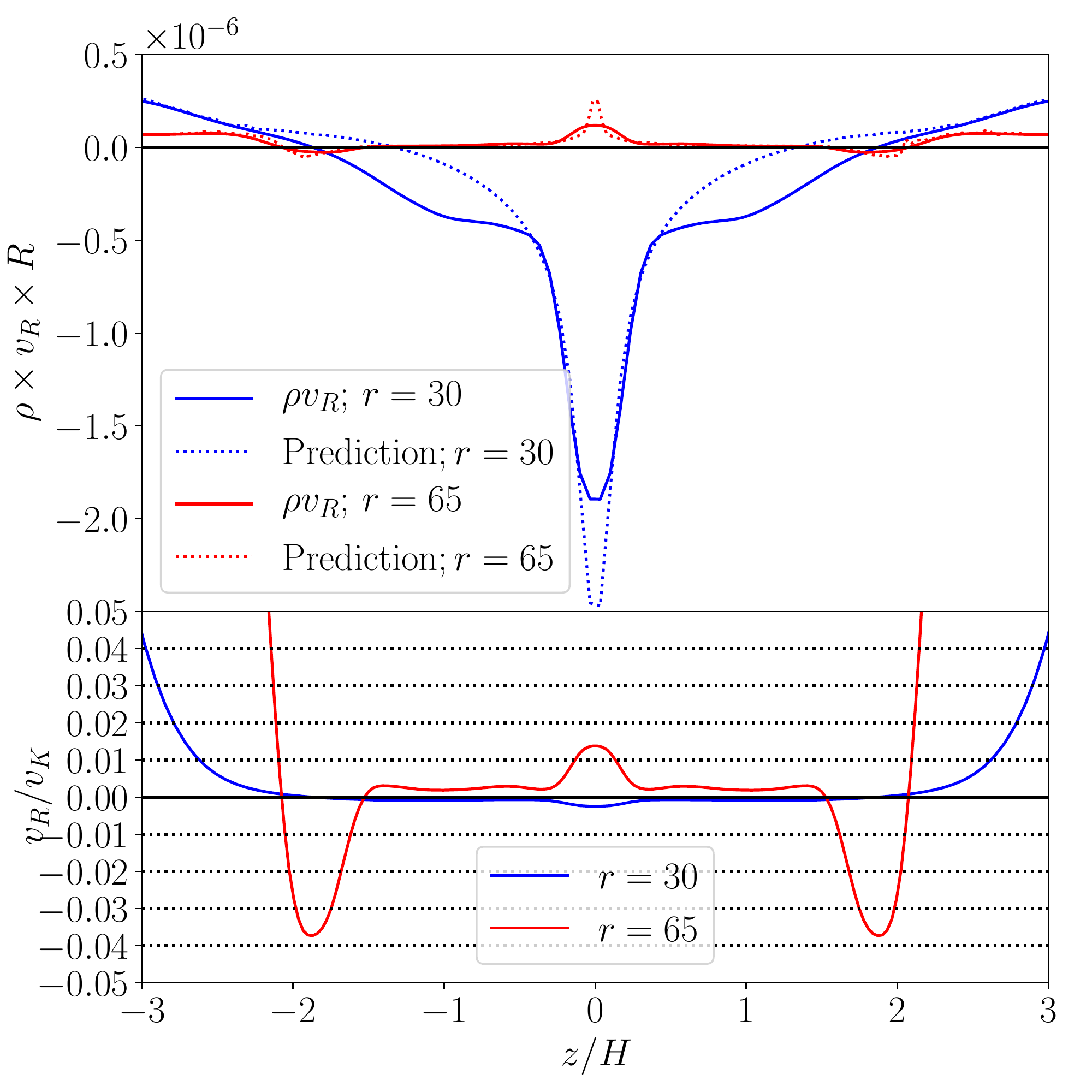}
    \caption{Vertical profiles of radial mass flux (upper) and radial velocity (lower) at two representative radii for Run \texttt{hi\_res}. The two radii are chosen to be within (blue, $R=35$) and outside of (red, $R=65$) the transition radius $R_t\approx50$). The data are averaged from $14000\Omega_0^{-1}$ to $20000\Omega_0^{-1}$. In the upper panel, measured values from data are shown in solid lines, while dotted lines show theoretical expectations from Equation~\eqref{eq:mflow}.}
    \label{fig:vcut}
\end{figure}

In Figure~\ref{fig:vcut}, we show the vertical profiles of radial mass flux $\rho v_R$ averaged from $14000\Omega_0^{-1}$ to $20000\Omega_0^{-1}$ as solid lines, as well as the
prediction based on the right hand side of Equation~\eqref{eq:mflow} as dotted lines. We choose $R=30$ as a representative location inside the transition point and $R=65$ as a representative location outside. We see that Equation (\ref{eq:mflow}) reasonably well reproduces the observed accreton/decretion flow structure within/beyond the loop center.
While there are deviations near the midplane, the agreement is reasonable given the approximate nature of the equation.

As discussed in Section \ref{subsec:mdot} and Appendix \ref{app:mlossrc}, the outflow mass flux beyond $\pm2H$ is connected to wind launched from beyond $r_c$.
As seen in Figure~\ref{fig:vcut}, the radial velocity increase rapidly beyond $\pm2H$ and can reach to $\sim0.65v_K$ at $\pm3H$ (not shown) and continues to be accelerated towards larger radii to exceed $v_K$. This flow is essentially unbound, similar to a standard disk wind.
This flow is analogous to but physically different from external photoevaporation, where recent 2D simulations of \citet{Haworth2019} illustrates that a substantial fraction of the mass loss arises from the surface of the outer disk. 
We will further discuss this analogy in Section \ref{ssec:evap}.

The bump in Figure \ref{fig:vcut} near $z=0$, on the other hand, best exhibits the magnetically-driven decretion flow. The flow speed is about $0.014v_K$ (remain bounded), which can give appreciable mass flux given higher gas density at the midplane. With a thickness of $\sim0.5H$, we find the outward mass flux it carries is about $\sim20\%$ ($\sim2\times10^{-6}$) of the total mass loss rate beyond $r_c$.
This is exactly analogous to conventional scenario of viscous spreading, implying that wind-driven evolution of PPDs can expand over time. 

\subsubsection{Density dispersion}

\begin{figure}[!htb]
    \centering
    \includegraphics[width=0.5\textwidth]{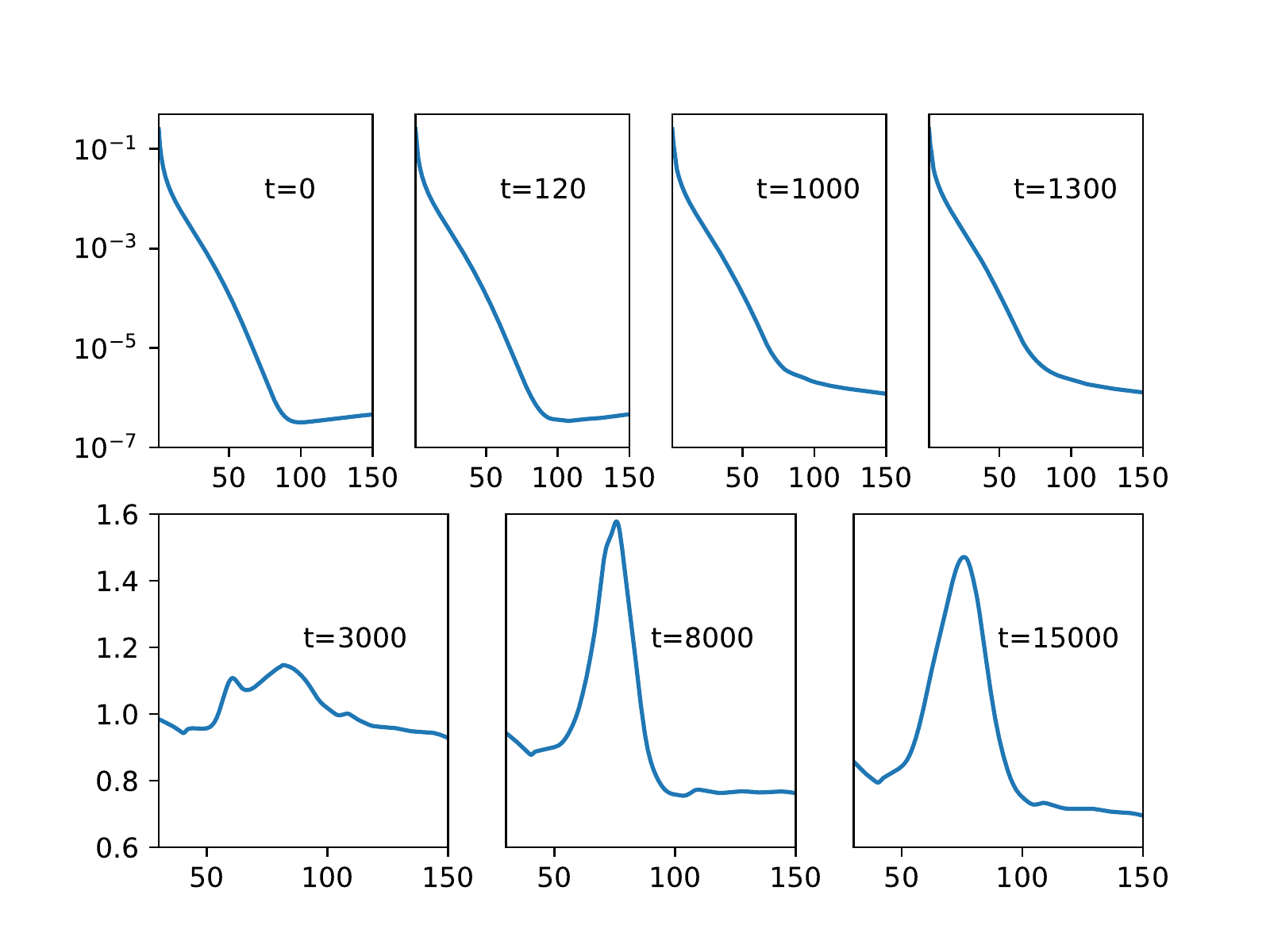}
    \caption{Upper panals: Column density distribution at $t=0,\,120,\,1000,\,1300\Omega_0^{-1}$, from left
    to right. 
    Lower panals: The ratio of the surface density to that of the reference time ($t=1300\Omega_0^{-1}$).
    From left to right we have $t=3000,\,8000,\,15000\,\Omega_0^{-1}$, respectively. 
 	We show only results beyond $r_c=30$ to focus on the part of the disk beyond the truncation radius, 
 	where our density replenishment scheme has little impact.}
    \label{fig:sigma}
\end{figure}

With mass outflowing beyond the loop center analogous to viscous spreading, 
here we analyze the dispersion of gas surface density.
In Figure \ref{fig:sigma}, we show a time sequence of the disk surface density profile (the integral of gas density along the 
$\theta$ direction).
We choose to plot only the part beyond the truncation radius ($r>r_c=30$), which is the focus of this work, and this region is
not affected by the density replenishment scheme inside $r_c$.

The upper four snapshots corresponds to stage 1 where the envelope is pushed away, establishing the density profile beyond the truncation radius. The last of them at time $t=1300$ (about one Keplerian period at $r_c=30$) sets a baseline for our further analysis.
After $t=1300\Omega_0^{-1}$, the column density profile evolves very slowly. 
To better visualize the evolution, we plot the ratio of the surface density profile to that at $t=1300\Omega_0^{-1}$ in linear scale.
We can see that the surface density inside of $r=50$ is gradually depleted due to the steady accretion flow. At the same time, an excess appears around $t=60-70$. This excess in surface density is comparable to the local surface density (on the orders of several tens of percent). The development of this excess is directly related to the decretion flow discussed earlier. Over time, the excess gradually flattens, as mass is further transported towards larger radii. Overall, the density dispersion we observe is indeed qualitatively analogous to expectations of viscous expansion. We with conduct a more quantitative comparison in Section \ref{ssec:spread}.

\subsection{Detailed analysis on flux evolution}
\label{subsec:ephi}
\begin{figure}[!htb]
    \centering
    \includegraphics[width=0.5\textwidth]{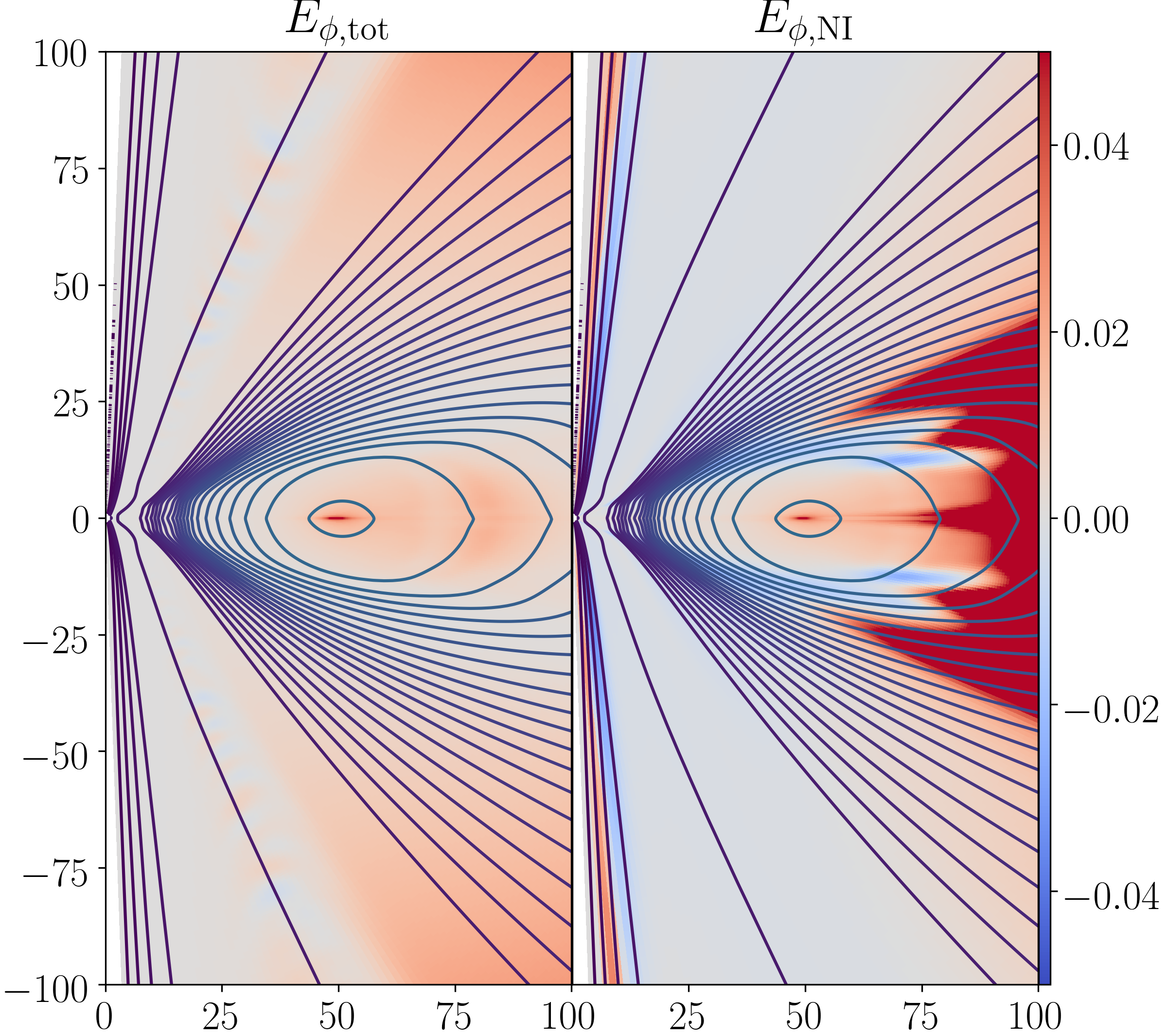}
    \caption{The $\phi$ component of the electric field, averaged over $t=14000-20000\Omega_0^{-1}$ for Run \texttt{hi\_res}. 
    The left panel shows the total $E_\phi$, whereas the right panel shows the contribution from non-ideal
    MHD effects. They are normalized by $v_K B_\mathrm{pol}$.}
    \label{fig:ephi}
\end{figure}

To better understand magnetic flux evolution in our simulations, 
in Figure~\ref{fig:ephi}, we show both the total $E_{\phi,\rm tot}$ (left panel) and its non-ideal contribution $E_{\phi,\rm NI}$, normalized by the product of the local Keplerian velocity $v_K$ and the poloidal magnetic field strength $B_{pol}\equiv\sqrt{B_r^2+B_\theta^2}$. Positive and negative $E_\phi$ are colored red and blue, correspondingly, and the color value thus marks the rate at which the field line is transported relative to $v_K$.

We first see from the left panel of Figure \ref{fig:ephi} that the total $E_\phi$ is largely positive throughout the system.
In the disk region, up to disk radii within the loop center where the poloidal magnetic field has a positive $z-$component, a positive $E_\phi$ directly translates to the loss of magnetic flux with outward flux transport.
Similarly, beyond the loop center where $B_z$ is negative, a positive $E_{\phi}$ leads to inward transport of magnetic flux. Combining the above, a positive $E_\phi$ essentially describes the shrinking of the field loop, and hence the dissipation and loss of magnetic flux.
More generally, a positive (negative) $E_\phi$ means the shrinking and dissipation of field loops if magnetic fields go clockwise (anti-clockwise) in the $(R,z)$ plane.
We see from the right panel of Figure \ref{fig:ephi} that in the bulk disk, as well as regions beyond the loop center, 
$E_{\phi,\rm tot}\approx E_{\phi,\rm NI}$. Therefore, the shrinking of magnetic flux loops are mainly due to the non-ideal MHD effects, i.e. ambipolar diffusion. In the following, we analyze flux transport more quantitatively.

\begin{figure}[!htb]
    \centering
    \includegraphics[width=0.5\textwidth]{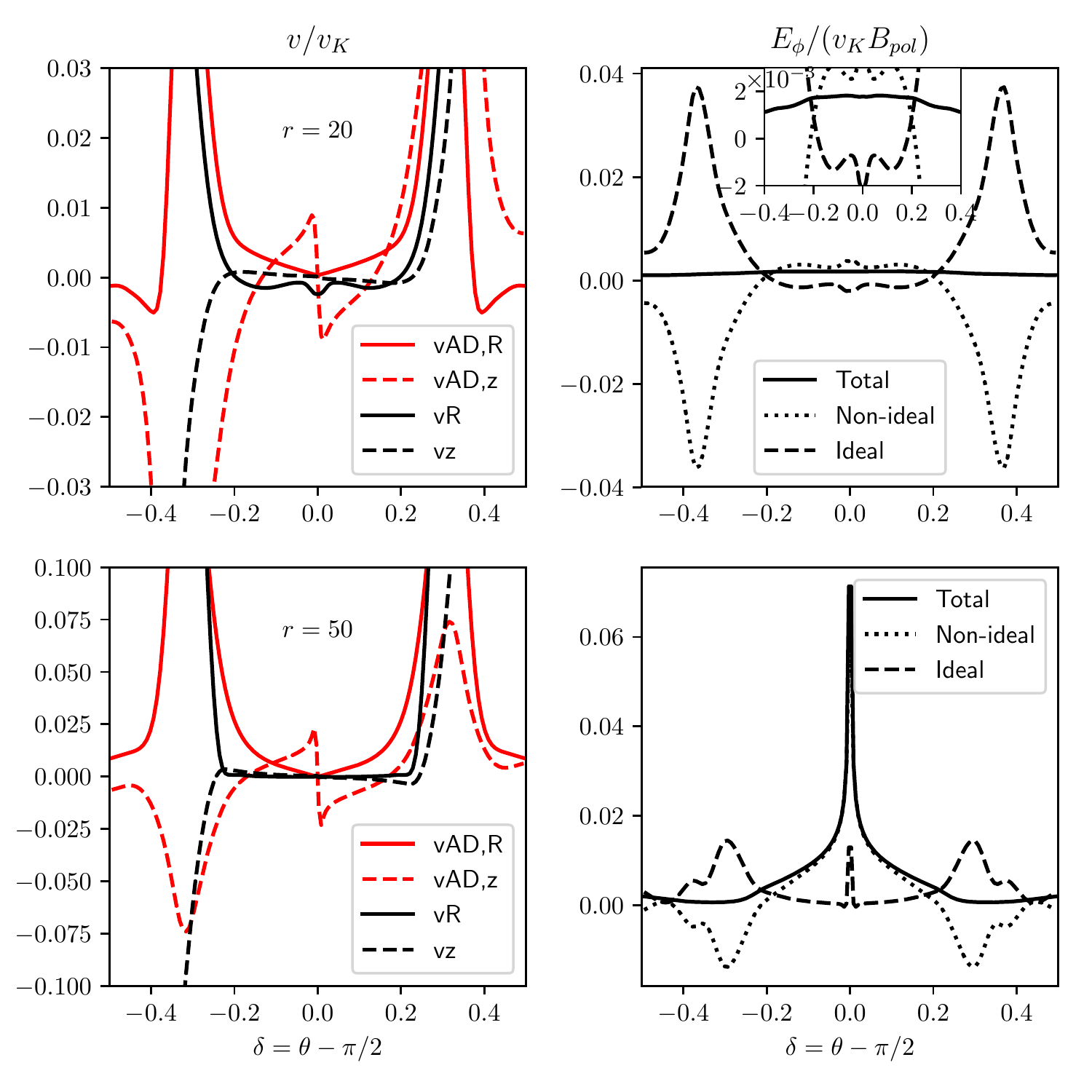}
    \caption{The $\theta$ profiles of various diagnostics that contribute to magnetic flux transport measured at spherical radius $r=20$ (upper row)
    and $r=50$ (lower row), averaged over $t=14000-20000\Omega_0^{-1}$ for Run \texttt{hi\_res}. 
    Left column: the ambipolar drift velocity $v_\mathrm{AD}$ and the flow velocity $v$, normalized by the local Keplerian velocity. 
    Shown are the (cylindrically) radial
    and vertical components of these velocities. Right column: the ideal and non-ideal terms in $E_\phi$, normalized by $v_KB_\mathrm{pol}$.
    The inset in the upper right panel is to enhance the visibility near the disk midplane.}
    \label{fig:veslice}
\end{figure}

In Figure~\ref{fig:veslice}, we show the $\theta$ profiles of mean gas velocity, ambipolar drift velocity (left panels) and the $E_\phi$ normalized by $v_KB_{pol}$ (right panels). The upper panels show the results at $r=20$, i.e. the disk region. The poloidal field is dominated by vertical component in the bulk disk, and the results are qualitatively similar to those reported by \cite{BS2017} in their Figure 5.  We can see that the $E_\phi$ curve is vertically-flat in the disk region, meaning a steady transport of magnetic fluxes across different heights. The exact value of flux transport rate, on the other hand, is smaller than that in \cite{BS2017} by roughly a factor of $3$. This difference likely results from the different treatment in the wind launching region, where we use $z_\mathrm{trans}=4$ for our fiducial run as opposed to $z_\mathrm{trans}=3$ in their work, with more discussions to follow in Section~\ref{sec:pars}.

The lower panels show the results at $r=50$, which is about the center of the magnetic flux loops. 
We can see that the gas poloidal velocity is close to zero near the midplane, thus transport is mainly mediated by ambipolar drift.
The ambipolar drift velocities show similar structures with those at $r=20$, largely due to the fact that the vertical profile of toroidal field remains similar. However, the poloidal field near this region becomes largely horizontal, rather than vertical. As a result, flux transport is mainly mediated by $v_{d,z}$ that brings radial field towards the loop center. This process is rapid, as exhibited in the bottom right panel, with a central spike in $E_\phi/B_{\rm pol}$. We note that if we simply plot $E_\phi$, the corresponding vertical profile becomes largely flat (i.e., steady dissipation of magnetic flux towards loop center). Therefore, rapid transport is compensated by the weak poloidal field near loop center. In other words, poloidal field is weakened or ``diluted", which can also be seen from the ``emptiness" of the constant-$\Phi$ contours near loop center from Figures \ref{fig:snapshots} and \ref{fig:ephi}.

Finally, we note that the flux transport rate at the innermost region is initially very fast, leading to rapid depletion of magnetic 
fluxes near the inner boundary, as can be seen in Figure~\ref{fig:snapshots}. 
This maybe physical because of the shorter dynamical timescale, but it is also affected by the inner boundary condition. 
This phenomenon is interesting on its own right, but also demands more realistic simulations (with better treatment of ionization and thermodynamics) for further study. In this paper, we restrict ourselves mainly on flux transport and dissipation beyond $r_c$.

\begin{figure*}[!htb]
    \centering
    \includegraphics[width=\textwidth]{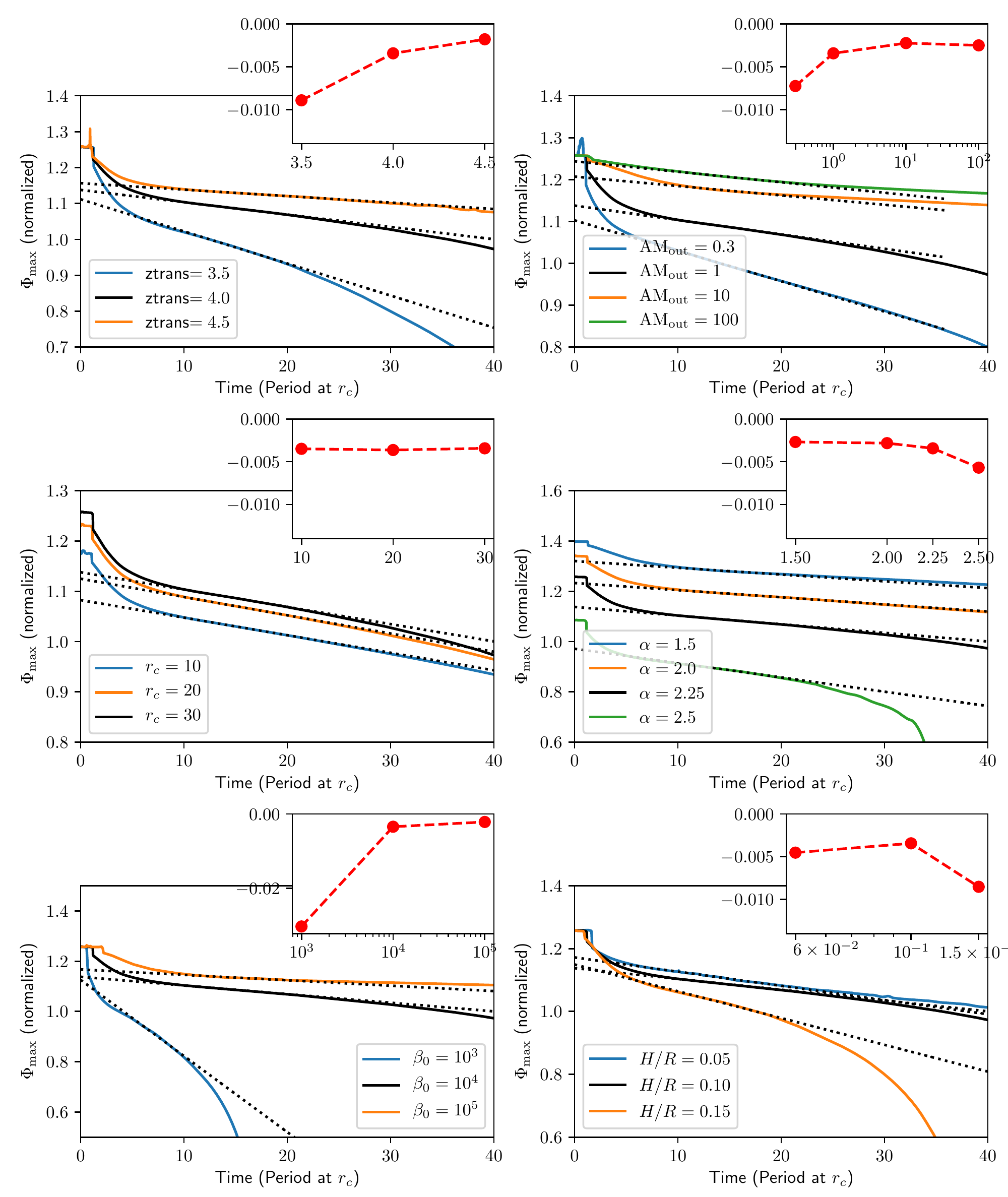}
    \caption{The same as Figure~\ref{fig:phimax} but for runs with varying parameters. The upper left panel 
    compares runs with different transition heights $z_{\rm trans}$. The upper right panel compares runs with different levels of AD coefficients at large radius. The middle left panel compares runs with different truncation radii $r_c$. The middle right panel compares
    runs with different power-law indices for the density profile. The lower left panel compares runs with different levels
    of magnetization (initial plasma $\beta_0$). The lower right panel compares runs with different aspect ratios of the 
    disk. In all panels, the fiducial run is plotted as solid black curves. The insets in all panels show the rate of magnetic flux loss
    $d\Phi_\mathrm{max}/dt$ versus the parameter of interest in that panel. The scale in all insets are the 
    same, except for the lower left panel comparing different runs with varying plasma $\beta_0$. }
    \label{fig:fluxes}
\end{figure*}

\section{Parameter Study}
\label{sec:pars}

In this section, we conduct a parameter study by varying one parameter at a time while fixing other parameters.
The sets of simulations include: 
\begin{enumerate}
\item The transition height for the AD Els\"{a}sser number ${Am}$, $z_\mathrm{trans}=3.5, 4.5$ (\texttt{zt3.5} and \texttt{zt4.5} in Table~\ref{tbl:runs}); 
\item The value of $Am$ beyond disk truncation radius, $\mathrm{AM}_\mathrm{out}=0.3, 10, 100$ (\texttt{AMout0.3}, \texttt{AMout10}, 
and \texttt{AMout100} in Table~\ref{tbl:runs});
\item The truncation radius $r_c=10, 20$ (\texttt{Rc10} and \texttt{Rc20} in Table~\ref{tbl:runs}), note that the maximum radius $r_\mathrm{max}$ is also modified accordingly to ensure $r_\mathrm{max}=10r_c$;
\item The power-law index for the disk density profile, $\alpha=1.5, 2.0, 2.5$ (\texttt{alpha1.5}, \texttt{alpha2.0}, 
and \texttt{alpha2.5} in Table~\ref{tbl:runs});
\item The initial disk magnetization $\beta_0=10^3,10^5$ (\texttt{beta1e3} and \texttt{beta1e5}, in Table~\ref{tbl:runs});
\item The disk aspect ratio $H/R=0.05, 0.15$ (\texttt{HoR05} and \texttt{HoR15}, in Table~\ref{tbl:runs}).
\end{enumerate}
Overall, we find that all of simulations show similar evolutionary trend as depicted in Figure \ref{fig:snapshots}, and we
discuss quantitative differences below.

The long-term system evolution is better characterized by the evolution of magnetic flux. We show in Figure~\ref{fig:fluxes} the
evolution of the $\Phi_\mathrm{max}$ for these simulations (normalized by the $\Phi_d$, from Equation \ref{eq:Phid}), grouped by different sets of parameters mentioned above.
We can see that almost all of the runs, upon entering stage 3, lose magnetic flux linearly over time.
We measure the slope of this linear relation, corresponding to the rate of flux losses, and show results in the insets, from which we can identify important trends in parameter dependence. Note that the scale for the $y$-axis is the same for all five insets except for that in the lower left panel comparing runs with different initial plasma $\beta_0$. 

First, we verify from simulations with different $r_c$ that the rate of flux transport, when measured in $P_c$ (Keplerian period at $r_c$), is independent of the specific choice of $r_c$. This result indicates that the overall rate of flux transport in disks is largely set by disk truncation radius, or simply disk size. Together with our other simulations with varying $\eta_O$, $T_{e0}$ and $\rho_{e0}$, they establish the robustness of our results against numerical parameters.

Among all parameters, the rate of flux transport depends most sensitively to $z_\mathrm{trans}$ and initial plasma $\beta_0$.
Stronger disk magnetization (smaller $\beta_0$) leads to faster rate of transport. This trend was seen in recent wind simulations of full disks \citep{BS2017,Lesur2021}, and it remains to hold in truncated disks. In particular, increasing magnetization from $\beta_0=10^4$ to $10^3$ leads to a dramatic increase in the rate by a factor of $\sim10$, whereas lower magnetization to $\beta_0=10^5$ leads to a modest reduction by $\sim40\%$. 
Lower/higher disk wind base (smaller $z_\mathrm{trans}$) enhances/reduces the loss rate of disk magnetic flux. A slight reduction of $z_{\rm trans}$ from our fiducial value of $4.0$ to $3.5$ increases the rate of flux loss by more than a factor of 2.5, whereas increasing $z_{\rm trans}$ to 4.5 reduces the transport rate by a factor of $\sim2$. This new trend that we identify here strengthens the importance to properly capture the physics of disk wind launching. It also helps explain the discrepancies in the absolute flux loss rate measured from the recent literature. For instance, the rate of flux loss in \cite{BS2017} is about a factor 3 faster than that obtained in our simulations, who adopted $z_\mathrm{trans}=3$.\footnote{On the other hand, \cite{Lesur2021} found much smaller transport rate though it is not straightforward to define an equivalent $z_{\rm trans}=3$ in his calculations: his $Am$ profile varies in a different manner as $\exp{(z/\lambda h)^4}$ with $\lambda=3$, and he assumes the gas to be isothermal, and hence there is no temperature transition.}.

The Els\"{a}sser number at large radius ($\mathrm{AM}_\mathrm{out}$) has moderate effects. In general, except for the last data point with $\mathrm{AM}_\mathrm{out}=100$, larger $Am$ number results in slower loss rate of magnetic flux. This result also exemplifies the influence of outer boundary conditions to the rate of flux transport.

By varying the density profile in the disk, parameterized by $\alpha$, we also vary the radial distribution of magnetic flux (since initial radial profile of net poloidal field has constant plasma $\beta_0$). Our parameter study shows that it has minor impact on the rate at which fluxes are lost. The main trend is that a steeper outward gradient of radial flux profile (larger $\alpha$) leads to faster rate of flux loss, which is in line with flux transport mediated by diffusion.
The rate of flux transport appears to be insensitive to the disk aspect ratio for $H/R\lesssim0.1$. This rate is modestly enhanced when the disk becomes thicker ($H/R=0.15$), although for $z_{\rm trans}=4$, the location of the wind base can no longer be considered as in a geometrically thin system.

\section{Discussion}
\label{sec:discussion}

There are three major aspects of our simulation results that merit further discussion, as we detail in Sections \ref{ssec:spread} - \ref{ssec:fluxtrans}. As a pilot study, our simulations are also subject to a number of caveats that will be discussed in Section \ref{ssec:caveats}.

\begin{figure*}[!htb]
    \centering
    \includegraphics[width=0.45\textwidth]{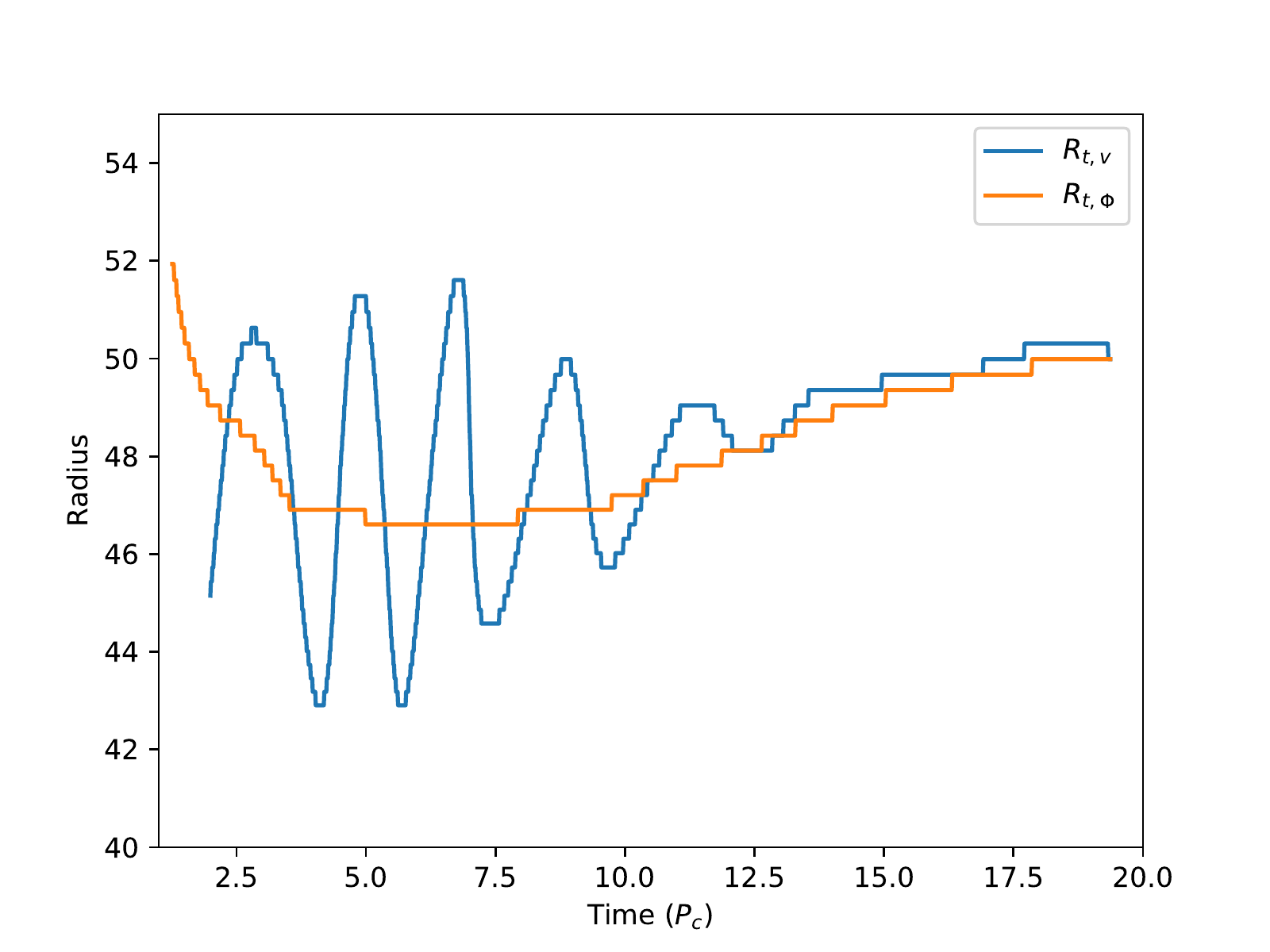}
    \includegraphics[width=0.45\textwidth]{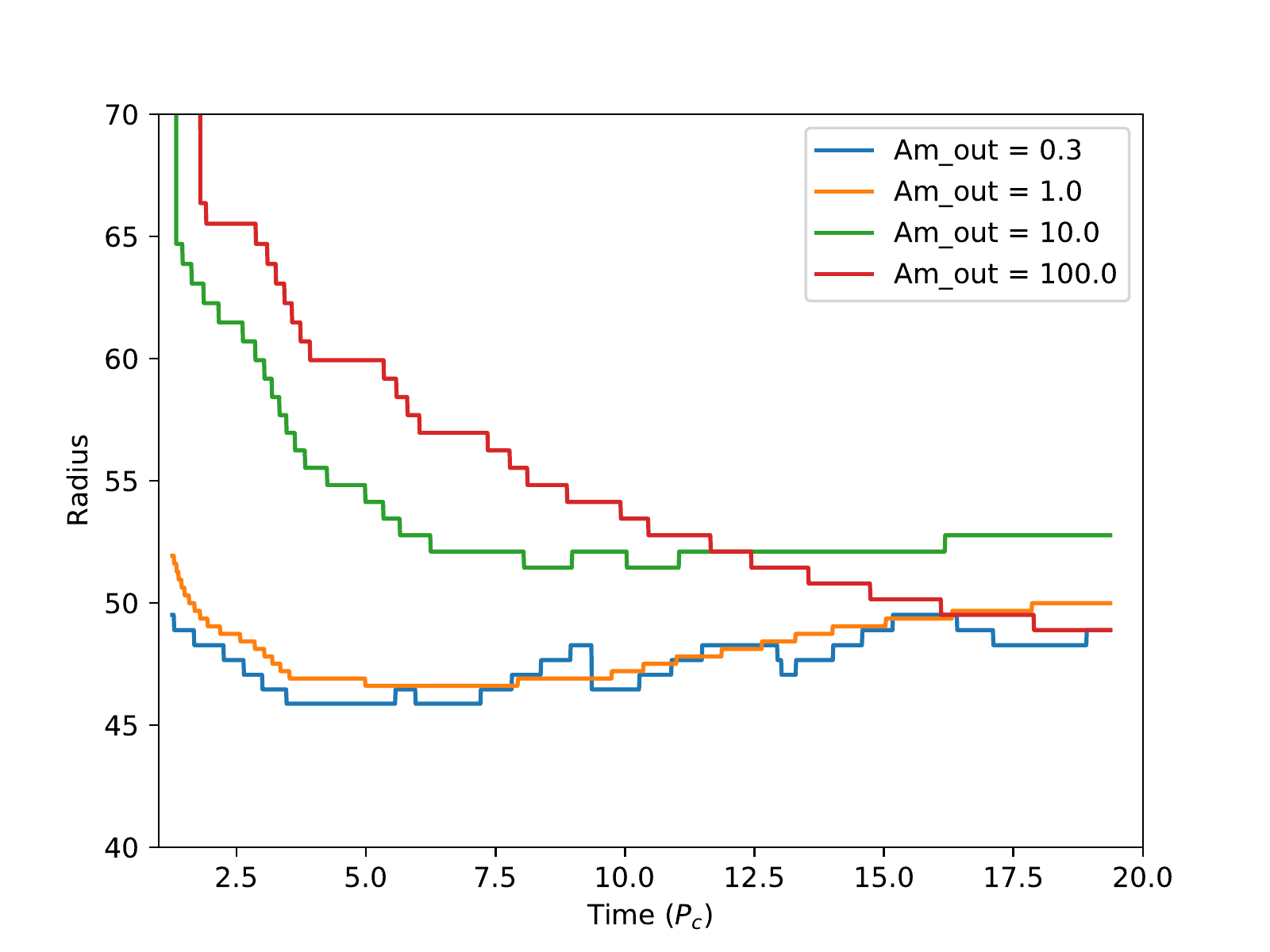}
    \caption{Left: The evolution of transition radius $R_t$ for Run \texttt{hi\_res}. Two transition radii are defined: $R_{t,v}$ is defined as the location where radial velocity changes its sign, and $R_{t,\Phi}$ is the location where the magnetic flux $\Phi$ is maximized. 
    The time is in unit of $P_c$ and the radius is in code unit.
    Right: The evolution of transition radius ($R_{t,\Phi}$) for runs with different $\mathrm{Am_{out}}$, taken from runs
    \texttt{AMout0.3}, \texttt{Fid}, \texttt{AMout10} and \texttt{AMout100}, respectively. }
    \label{fig:rt}
\end{figure*}

\subsection{Analogy with viscous spreading}\label{ssec:spread}

Canonically, models of long-term disk evolution assume outward transport of disk angular momentum transport mediated by viscosity. As a result, most materials accrete, transferring their angular momentum to the outer disk, which expands. There is a transition radius separating accreting and expanding region, and this transition radius itself increases with time. Here we compare the outcome of our simulation results to that of the viscous evolution model to show the analogy and potential differences.

\subsubsection[]{Expectations from viscous spreading}

For a disk with viscosity profile of $\nu\propto R^{\gamma}$, the viscous evolution equation of surface density has a similarity solution, with transition radius evolving as \citep{LP1974,Hartmann1998}:
\begin{equation}
R_t=R_1\left[\frac{T}{2(2-\gamma)}\right]^{\frac{1}{2-\gamma}}\ ,
\label{eq:Rt_alpha}
\end{equation}
where $R_1$ is a radial scale factor, $T=t/t_s+1$ is a nondimensional time with $t_s=(R_1^2/\nu_1)/3/(2-\gamma)^2$ being the viscous scaling time, and $\nu_1$ is the viscosity at $R_1$.
Taking the derivative of Equation~\eqref{eq:Rt_alpha}, one gets the expansion rate of the transition radius as:
\begin{equation}
\frac{dR_t}{dt} = \frac{R_1}{t_s} \frac{1}{2-\gamma}\left[\frac{T}{2(2-\gamma)}\right]^{\frac{\gamma-1}{2-\gamma}}\ .
\end{equation}
For simplicity, we will assume $\gamma=1$. Results assuming other $\gamma$ are generally not very different. In this case, we have $dR_t/dt = R_1/t_s$, with $t_s=R_1^2/(3\nu_1)$. For a typical $\alpha$ disk with $\nu_1 = \alpha c_s H$, we find:
\begin{equation}
\frac{dR_t}{dt} = 3\alpha \left(\frac{H}{R}\right)^2 v_K\ .
\end{equation}

This rate of spreading is directly linked to the radial flow velocity.
Using Equations~(17) and (21) from \cite{Hartmann1998}, one can derive the radial profile of the radial velocity, for $\gamma=1$ case, as:
\begin{equation}
v_R = \frac{3}{2}\frac{\nu_1}{R_1}\left(\frac{R}{R_t}-1\right)
=\frac{3}{2}\alpha \left(\frac{H}{R}\right)^2 v_K\left(\frac{R}{R_t}-1\right)\ .
\end{equation}

\subsubsection[]{Expansion of transition radius $R_t$}

A magnetized disk wind is expected, conventionally, to extract angular momentum from the disk, driving the entire disk to accrete. Our simulations show that this is not the case due to the formation of magnetic field loops beyond truncation radius. Instead, there exists a similar transition radius, which separates accretion and decretion flows analogous to viscous spreading. We have also seen that the mass flux beyond the transition radius is a sizable fraction of the accretion flow. 
It remains to examine how the transition radius itself evolves.

In the left panel of Figure~\ref{fig:rt}, we plot the evolution of the transition radius $R_{t,v}$, defined by the location where radial velocity changes sign in the midplane, and the evolution of the location where the magnetic flux $\Phi$ is maximized ($R_{t,\Phi}$, i.e., loop center) from run \texttt{hi\_res}. 
We see that $R_{t,\Phi}$ closely tracks the center of the magnetic flux loops. Following some relaxation processes in stage 2 where the field loop center oscillates, the two closely match each other in stage 3 (which is one main reason we choose $t\approx 14P_c$ as the starting time of stage 3).
Interestingly, in stage 3, we observe that the transition radius also
moves outward. This is again analogous to the scenario of viscous spreading.

Fitting the linear part in the left panel of the Figure~\ref{fig:rt}, we get $dR_t/dt\sim 1.4\times 10^{-3} v_K$, where the Keplerian velocity was taken at $r=50$. If this expansion rate were due to viscous spreading, we find an effective $\alpha \approx 0.047$. 

By examining our other simulation runs, we find that the evolution trend of $R_{t,\Phi}$ is mainly sensitive to the value of $Am_{\rm out}$, reflecting the ionization fraction in the envelope. 
The rate of $dR_{t,\Phi}/dt$ can be modified by other parameters at some modest level but without sign change. Here we will focus on the role of $Am_{\rm out}$, and the results are shown in the right panel of Figure~\ref{fig:rt}.
We can see that runs with relatively small $\mathrm{Am_{out}}\lesssim1-10$ (and hence lower ionization level) tend to have $R_{t,\Phi}$ expand over time, whereas runs with large $\mathrm{Am_{out}}=100$ (and hence higher ionization level) tend to have $R_{t,\Phi}$ decrease with time. 

Given the importance of ${\rm Am_{out}}$, here we give a rough estimate of the correspondence between the value of Am and the ionization fraction. At the transition radius of $r=50$, we find $\rho=8.5\times 10^{-6}$ in code units, averaged over $10000-20000\Omega_0^{-1}$
for the \texttt{hi\_res} run. Using the same unit conversion method discussed in Section~\ref{ssec:diags}, we find $\rho_0=9.2\times 10^{-12}\rm\, g/cm^3$. Assuming HCO+ as the dominant ion with $\left<\sigma v\right>\approx 2\times 10^{-9}\rm\, cm^3/s$, we obtain the ionization fraction $x_e\approx 3\times 10^{-9}\mathrm{Am}$.
The real ionization fraction in the outer regions of PPDs is largely uncertain. Calculations from \cite{Cleeves2015} for the IM Lup disk suggests that $x_e$ to be between $10^{-11}$ to $10^{-9.5}$, translating to $Am=0.004\sim 0.1$, thus the disk evolution likely follows an overall expanding path. Note that they have adopted a substantially reduced cosmic-ray ionization rate with $\zeta_\mathrm{CR}<10^{-19}\, \rm s^{-1}$, whereas the value of $Am$ could be brought to order unity for more standard ionization rates.
Nevertheless, the exact ionization rate and hence ionization level is still highly uncertain and likely depend on external environment, and higher $Am$ value is likely when the disk is in the vicinity of some massive stars (see next subsection). In such cases, despite having a decretion flow beyond $R_{t,\Phi}$, the disk itself still likely shrinks over time due to the contraction of $R_{t,\Phi}$.

\subsubsection[]{Radial flow speed beyond $R_t$}\label{sssec:vrflow}

To compare our radial velocity with the 1D viscous model, we conduct density-weighted average on 
radial velocity vertically between $-3H$ to $3H$ at $R=65r_0$. We get $\overline{v_R}(R=65r_0)=8.7\times 10^{-3}v_K$. Taking $R_t\sim 49.7r_0$, we find that $\alpha = 1.88$ if this radial flow velocity were due to viscous spreading. This value is apparently too large for any reasonable viscous disk models, partly because the mass outflow is dominated by high-velocity wind flows (which is analogous to external photoevaporation). If we limit ourselves to only the mass flux in the midplane region between $\theta=\pi/2\pm 0.05$ but still average over the entire disk column density, we find $\overline{v_{R,\rm mid}}=2.7\times 10^{-3}$, which translates to $\alpha=0.58$. This is still more than one order of magnitude larger than the value inferred from $dR_t/dt$.

The large apparent equivalent $\alpha$ value here reflects the highly-efficient nature of magnetic torques exerted to the outer disk. As it is analogous to (but reversed) the wind-torque that transports angular momentum vertically, it is more efficient by a factor of $\sim R/H$ than the equivalent viscous torque assuming similar field strengths (e.g., \citealt{Wardle2007,Bai2009}). 

\begin{figure}
    \centering
    \includegraphics[width=0.5\textwidth]{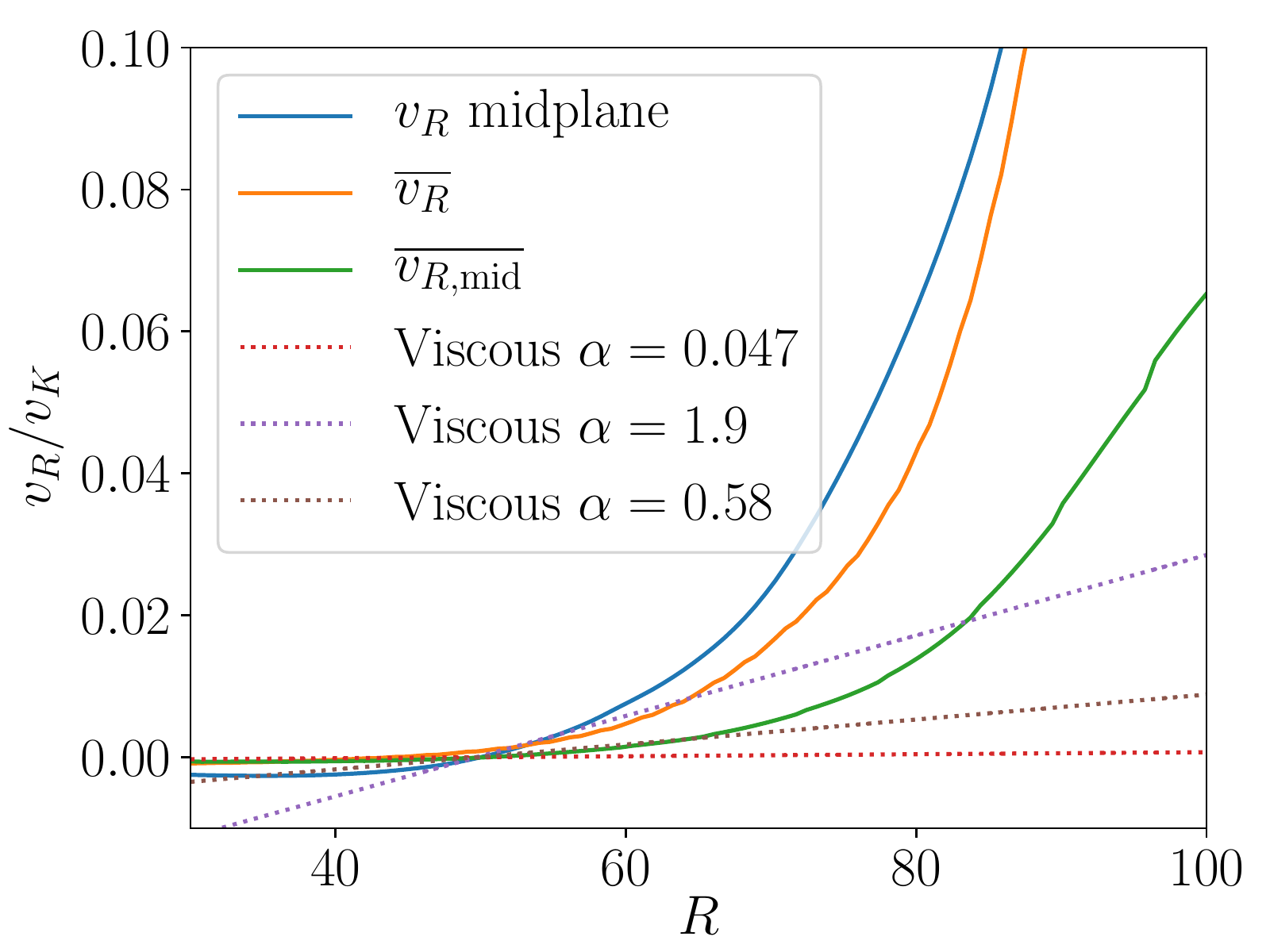}
    \caption{The radial velocity profiles for the fiducial run, averaged between $t=14000\Omega_0^{-1}$ and $20000\Omega_0^{-1}$. See the Section \ref{sssec:vrflow} for the definition of $\overline{v_R}$ and $\overline{v_{R,\rm mid}}$. Also plotted are the radial velocity profile for a viscous disk with different $\alpha$. We have assumed density power law index $\gamma=1$ for simplicity \citep{Hartmann1998}.}
    \label{fig:vr}
\end{figure}

In Figure~\ref{fig:vr}, we show the three different radial velocity profiles, $v_R$ at midplane, $\overline{v_R}$, and $\overline{v_{R,\rm mid}}$ defined above. Also plotted are the radial velocity profiles for a viscous spreading disk with $\alpha=0.047$, $1.9$, and $0.58$. The difference in radial velocity profiles between disk wind driven flow and viscous spreading is clearly shown. We see that mean radial velocity not only requires a large effective $\alpha$, they also increase rapidly over radius than that of a typical viscous model. 
This may offer clues to disentangle the decretion flow found in wind-driven disk evolution from that of standard viscous evolution, although more systematic modeling is beyond the scope of this work.

Based on the discussions above, we can tentatively conclude that the overall gas disk size, under typical conditions, likely expand over time, and this is accompanied by a decretion flow. The overall scenario qualitatively matches all aspects of viscous spreading in canonical models of disk evolution, except that the mass flux of the decretion flow well exceeds the viscous counterpart. Given that our simulation settings are highly simplified, we are not yet in a position to develop a long-term disk evolutionary model for quantitative comparison. On the other hand, our results suggest that it is premature to distinguish the mechanisms that drive disk angular momentum transport by searching for signature of disk expansion. By comparing gas disk sizes at different disk evolution stages, there is tentative evidence of a trend of larger disk size \citep{Najita2018,Trapman2020}. We therefore argue that such signatures are equally likely to be the outcome of wind-driven disk evolution, based on our findings.

\subsection[]{Relation to external photoevaporation}\label{ssec:evap}

External photoevaporation has been considered as another important mechanism that drives disk dispersal. This occurs when stars are strongly irradiated by the UV photons from nearby massive stars, which can be a common situation as stars form in clusters. This leads to strong mass loss beyond the gravitational radius $r_g$, where gas is heated such that its sound speed approaches the escape velocity \citep{Hollenbach1994,Johnstone1998}, while later studies found that it can be relaxed to
small radii of $\gtrsim 0.1-0.2r_g$ \citep{Adams2004,Facchini2016}. It is anticipated that 
external photoevaporation can be important in disk dispersal even for very mild environment with $G_\mathrm{FUV}=10-30\rm\, G_0$ \citep{Facchini2016,Haworth2018}, where $G_0$ is the Habing unit of UV radiation \citep{Habing1968} representing interstellar radiation field, though most previous studies were based on simple 1D calculations. More recent 2D hydrodynamic simulations with extensive photodissociation physics have found modest increase of mass loss rate \citep{Haworth2019}.

In our fiducial run, the temperature of the ambient medium is set to constant at $T_{e0}=0.001 (GM/r_0)$. The gravitational radius is roughly $1000 r_0$, which is well beyond our simulation domain and about $20$ times our transition radius. Therefore, external photoevaporation is not expected to operate. We have seen that we still have strong mass outflow emanating from beyond the truncation radius. These flows can be interpreted as an extension of the disk winds, but launched beyond truncation radius. They reside above a couple disk scale heights, typically having much higher speeds than the midplane decretion flow, but their flow properties are strongly affected by non-ideal MHD effects (e.g., gas does not flow along field lines).
Such flows are exactly analogous to the external photoevaporation yet it is mediated by magnetic fields. It is conceivable that in reality, both magnetic and thermal effects matter, similar to the picture of magneto-thermal disk winds \citep{Bai2016}, but the non-ideal MHD nature of this outer region may preclude developing a semi-analytical theory.

It is interesting to note that strong external photoevaporation usually implies higher ionization fraction in the disk surroundings, or larger $Am_{\rm out}$. From our results, we see that transition radius $R_t$ will shrink over time, suggestive of shrinking disk size. This adds to another layer of analogy with the picture of external photoevaporation, which also reduces disk size despite of viscous spreading in action 
\citep{Clarke2007,Anderson2013}.
Nevertheless, our rather rough treatment of thermodynamics restrain us from drawing more quantitative conclusions, and we will defer more detailed comparisons for future studies.

\subsection{Global transport of magnetic fluxes}\label{ssec:fluxtrans}

The transport of magnetic flux governs long-term disk evolution, and is intrinsically a global problem. Our work fills an important gap in the study of flux transport by providing more realistic outer boundary conditions. We have shown that the formation of magnetic flux loops is inevitable beyond truncation radius, and flux transport at global scale is governed by the shrinking of such field loops. We have identified several interesting trends as discussed earlier.
These trends indicate that conditions both in the disk (e.g., $z_{\rm trans}$) and beyond the truncation radius (e.g., ${\rm Am_{out}}$) matter for global rate of flux transport, thus exemplifies the notion that flux transport is intrinsically a global problem. 

This situation is very different from conventional theoretical models of flux transport, where the disk is typically assumed to be infinitely extended radially to permit radially-self-similar-type solutions. Under such assumptions, recent works start to incorporate more detailed disk vertical structure, as well as the role of disk winds \citep{Leung2019,Lesur2021}. We anticipate that such approaches may still be applicable well within the truncation radius $r_c$, but new theories are needed for regions beyond disk truncation to determine the global rate of flux transport. 

While developing a new theory of flux transport is beyond the scope of this work, we here make a simple attempt by comparing the rate of flux loss and the characteristic rate of magnetic dissipation by ambipolar diffusion. When considering a characteristic length scale of $R_p$ at loop center, the latter is given by
\begin{equation}
    \frac{1}{\Phi_d}\frac{d\Phi_{\rm max}}{dt}\bigg|_{\rm AD}\sim\frac{\eta_A}{R_p^2}
    =\frac{4\pi(H/R)^2}{{\rm Am}\ \beta}P_c^{-1}\ .
\end{equation}
When considering $H/R\sim0.1$, $\beta\sim100$ in the midplane, $Am\sim1$, we obtain a rate of $\sim1.3\times10^{-3}P_c^{-1}$, which is smaller but comparable to the rate obtained in our simulations of $\sim3.5\times10^{-3}P_c^{-1}$. However, if we replace $R_p^2$ by $H^2$ at radius $R_p$, relevant to dissipation near loop center, the rate becomes much higher and is inconsistent with our measured flux dissipation rate. This disparity is already hinted in Section \ref{subsec:ephi}, where we find rapid dragging of poloidal field towards the loop center from above and below thanks to ambipolar drift, leading to a dilution of field lines. If flux transport is mainly mediated by diffusion, we see that the characteristic length scale is better chosen to be close to $R_c$. Nevertheless, while the scenario of diffusive transport is consistent with the trend that enhancing AD in the otuer disk (smaller ${\rm Am_{out}}$) leads to faster transport, this trend does not scale linearly as ${\rm Am_{out}}$ based on our simulations. It is likely that a combination of diffusive and advective transport is needed to account for the observed results, which contribute differently at different radii and heights, as seen in Figure \ref{fig:veslice}.

Using the same conversion method as discussed in Section.~\ref{ssec:diags}, we have $r_0=3$ AU, which corresponds to a transition radius of $150$ AU and a truncation radius of $90$ AU. The Keplerian period at $r_c$ is thus $P_c=850$ yr. 
The rate of flux dissipation measured above with $d\Phi_\mathrm{max}/dt\sim 3.5\times 10^{-3}\Phi_d/Pc$ can be 
directly translate to a flux dissipation timescale of $286P_c=0.24\rm\, $Myr. This is a lot shorter than typical disk lifetime of $2-3$ Myr \citep{Haisch2001,Mamajek2009}. On the one hand, owing to a number of caveats (see next subsection), and similar to previous work \citep{BS2017}, we here mainly focus on the trends, and do not anticipate the absolute rate of flux transport to be realistic. On the other hand, if the real rate of flux loss in disks is not far from what we have measured, the rapid flux loss might imply an early transition from magnetically-driven evolution to primarily hydrodynamically-driven evolution, and would have major implications to the long-term disk evolution theory and planet formation.

\subsection{Caveats}\label{ssec:caveats}

As a first study, we made several simplifications to the problem to ease the computational effort. While we expect our simulations to capture the most important physics, some caution should be exercised as we discuss below.

Firstly, our treatment of disk chemistry and thermodynamics is highly simplified, leading to several free parameters. Some of the parameters that we vary systematically (such as $z_{\rm trans}$, ${\rm Am_{out}}$), can be considered as reasonable proxy for physical conditions around the disk, while it is less certain about a few other fixed parameters, especially with respect to the temperature prescription, and the Am profile.
In reality, temperature in the outer disk is mainly determined by irradiation both from the central object and 
from the external sources. Coupled with it involves photo-chemistry and ionization chemistry, which determine cooling in the disk atmosphere as well as the ionization level in the system (e.g., \citealt{Haworth2016, WangBG2019, Gressel2020}). We thus anticipate that our calculations are subject to such systematics that demand more realistic calculations incorporating such microphysical processes, but the trends identified in our work are likely robust.

Another important caveat is that our simulations are two-dimensional assuming axisymmetry. In particular, given the level of Am value in the disk region and beyond, the disk is expected to be MRI turbulent, as has recently been demonstrated in full 3D global simulations \citep{CuiBai2021}. Besides being turbulent, another main difference is that unlike our 2D case where toroidal field changes sign exactly at the midplane, toroidal field changes sign at random heights which dynamically evolve, which affects the global flow structures. Moreover, magnetic flux is found to concentrate to quasi-axisymmetric sheets, which may reduce the global rate of flux transport. It is yet to be seen how the MRI turbulence beyond truncation radius affect the dissipation of large-scale poloidal magnetic flux loops encircling the outer disk, which we leave for future work.

\section{Conclusions}
\label{sec:conclusion}

In this work, we performed the first global non-ideal MHD simulations of PPDs with outer truncation. Our simulations follow the recently established paradigm of disk evolution driven by magnetized disk winds, yet compliment previous studies by incorporating more realistic boundary conditions to mimic disk outer truncation. As the disk outer regions typically contain most of the disk mass and largely govern the disk evolutionary timescale, we aim to clarify issues related to the long-term evolution of PPDs and their interplay with interstellar environment. Given that wind-driven accretion is largely controlled by the amount of poloidal magnetic flux threading disks, we pay special attention to the global transport of magnetic flux. 

Starting from hour-glass shaped field threading the disk, we find  that as the disk launches a magnetized disk wind, which quickly pushes away the surrounding materials, yet because loss of disk pressure support, poloidal field lines collapse beyond truncation radius, reconnect, and form field loops. The loops gradually relax towards a quasi-steady state that is largely independent of initial conditions, and it is such state that best represents realistic PPDs.
We have conducted a large parameter survey on top of a detailed study under a fiducial set of parameters, and our main findings are as follows.

\begin{itemize}
    \item The center of the poloidal field loop encircling the outer disk separates the disk into accretion (inside loop center) and decretion (outside loop center) flow regions, as a natural consequence of magnetic field geometry around the loop.
    \item Unless the disk outer region is well ionized, the loop center migrates outward over time, and the overall outer disk evolution is directly analogous to viscous spreading even in the absence of viscosity.
    \item Launching of disk winds extends to beyond truncation radius despite non-ideal coupling of gas with magnetic field, leading to significant mass loss analogous to external photoevaporation but without thermal driving.
    \item Global evolution of poloidal magnetic flux is largely governed by dissipation within poloidal field loops. The rate of dissipation is sensitive to both disk conditions and outer boundary conditions from disk truncation.
\end{itemize}

Our results imply that in the wind-driven accretion scenario, disk sizes will likely grow, rather than shrink as conventionally assumed, in time, similar to the scenario of viscously-driven disk accretion. The results are qualitatively consistent with recent observational inference of larger disk sizes towards older disks, and such observational evidence is insufficient to distinguish between the two driving mechanisms of disk angular momentum transport. We do note, however, that the gas velocity in the decretion flow is much faster than the viscous counterpart, which may offer a way to distinguish from the viscous evolution scenario.

Our results also imply that mass loss through disk outer boundary does not necessarily be thermally-driven, but can equally well be magnetically-driven. In reality, it is likely that both thermal and magnetic effects matter, which jointly drive the disk outflows in a manner analogous to magneto-thermal disk winds.

Our study highlights the importance of outer boundary conditions, besides disk microphysics, in determining the global magnetic flux transport in disks, calling for improved, more global theory of flux transport. As flux evolution and disk evolution are coupled, a global picture of disk evolution would be incomplete without a solid understanding of magnetic flux transport.

Finally, our simulations are in 2D, with simplified treatment of thermodynamics and chemistry. While we anticipate the trends identified in our study are robust, extensions to 3D as well as incorporating more realistic radiative physics and chemistry are needed to yield more quantitative and realistic results.

\section*{Acknowledgements}

We thank the anonymous referee for a prompt report with useful suggestions, and Can Cui for kindly sharing her problem setup with us and for fruitful discussion. This work is supported by the National Key R\&D Program of China (No.2019YFA0405100). Numerical simulations are conducted on TianHe-1 (A) at National Supercomputer Center in Tianjin, China, and on the Orion cluster at Department of Astronomy, Tsinghua University.

\appendix

\section[]{Composition of mass outflow beyond truncation radius: Further analysis}\label{app:mlossrc}

\begin{figure}[!htb]
    \centering
    \includegraphics[width=0.5\textwidth]{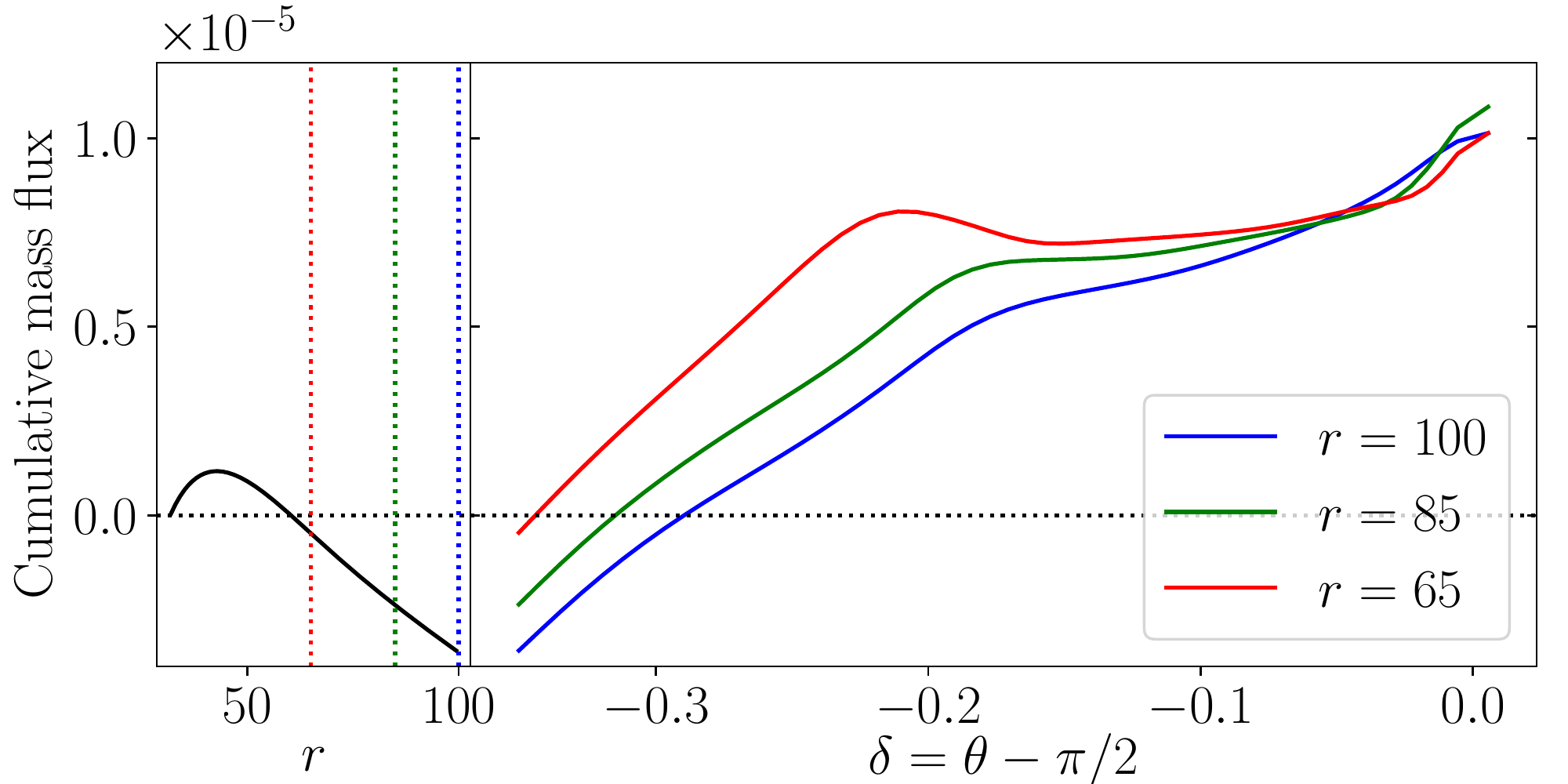}
    \caption{Left: the cumulative mass flux through the upper and lower contour shown in Figure~\ref{fig:mdot} between $r_c$ and $r$. 
    Right: the cumulative mass flux through the outer arc at $r=65$, $85$, and $100$, correspondingly. 
    See text for more details. The data are taken from Run \texttt{hi\_res} after averaging between
    $t=14000$ to $20000\Omega_0^{-1}$.} 
    \label{fig:cumflux}
\end{figure}

To further analyze the origin the mass flow beyond $r_c$ more quantitatively, we integrate the mass flux through the constant $\delta\equiv\theta-\pi/2$ part of the contour shown in Figure~\ref{fig:mdot}. The cumulative flux as a function of $r$ is shown on the left panel of Figure~\ref{fig:cumflux}. Note that the mass flow from both the upper and lower sides of the disk are added.
We can see that the cumulative mass flux first increases, then decreases starting from $r\sim 45$, indicating gas is flowing into the region enclosed by the contour.
At the end of the constant $\delta$ lines, we continue the integration to calculate the cumulative mass flux along the constant-$r$ arc, 
from $\delta=\pm 0.35$ down to $\delta=0$. Again, the mass flux from the upper and the lower side are added together.
We also do the same integration changing the constant $r$ arc to $r=65$ and $r=85$. Note that the starting point of the curves are changed depending on the amount of mass flux across the constant $\delta$ contour inside of the radius of interest.
The results for all three radii are plotted in the right panel of Figure~\ref{fig:cumflux}. The intersection of the three vertical dotted lines with the black line in the left panel correspond to the starting point of the three curves on the right with matching colors.

The overall cumulative mass flux along the constant-$\delta$ line and constant-$r$ arc indicates that first, there is wind launching beyond $r_c$. The orientation of the wind is at they first penetrate out of the constant-$\delta$ lines, but then bend downwards (due to disk truncation) to penetrate into these contour lines. They then exit from the outer constant-$r$ arc, leading to cancellation (when $\delta$ reaches $\sim\pm0.3$). Further increase of mass flux in the arc section indicates net mass loss beyond $r_c$. From the arc at $r=65-100$, we see a steady increase up to $\delta=\pm0.2$. We interpret this steady increase as a continuation of disk wind from beyond $r_c$. From the streamline plot in Figure \ref{fig:streamline}, we see these flows are launched typically within transition radius $r_t$). They are also heavily modified by non-ideal MHD processes (as opposed to standard wind scenario) as gas and magnetic fields are not well coupled. From Figure \ref{fig:cumflux}, we can roughly infer that this part of the outflow accounts for about $\sim70\%$ of the quoted total mass loss rate beyond $r_c$.
To some extent, this flow is somewhat analogous to mass outflow resulting from external photoevaporation.

The cumulative mass flux enters a plateau roughly between $\delta=\pm0.2$ and $\pm0.05$, where it increases very slowly.
Then, very close to the midplane within $\delta=\pm0.05$, the outward mass flux experience another rapid increase, and by reaching $\delta=0$, the total cumulative mass flux reaches the value quoted in Section \ref{ssec:mlossrc}. This outward mass flux near $\delta=0$ corresponds to the decretion flow discussed in detail in Section \ref{subsec:mflow}, and is analogous to viscous spreading. More quantitatively, we find that this outward mass flux corresponds to about $20\%$ of quoted total mass loss rate beyond $r_c$.

\end{document}